\documentclass[strctabstract]{aa}  

\usepackage{txfonts}
\usepackage{natbib}
\bibpunct{(}{)}{;}{a}{}{,}
\usepackage{graphicx}
\begin{document}
\title{Radio emission from the massive stars in the Galactic Super Star Cluster Westerlund 1}
\author{S.M.~Dougherty\inst{1,2}
\and J.S.~Clark\inst{3}
\and I. Negueruela\inst{4}
\and T. Johnson\inst{1,5}
\and J.M. Chapman\inst{6}}

\institute{$^1$National Research Council of Canada, Herzberg Institute for
Astrophysics, Dominion Radio Astrophysical Observatory, P.O. Box 248,
Penticton, British Columbia V2A 6J9, Canada \\
$^2$ Institute for Space Imaging Science, University of Calgary, 2500
University Dr. NW., Calgary, Alberta, T2N 1N4, Canada \\
$^3$ Department of Physics and Astronomy, The Open 
University, Walton Hall, Milton Keynes, MK7 6AA, UK\\ 
$^4$ Dpto. de F\'{\i}sica, Ingenier\'{\i}a de Sistemas y Teor\'{\i}a de la
Se\~{n}al, Universidad de Alicante, Apdo. 99, E03080 Alicante, Spain\\
$^5$ Department of Physics and Astronomy, University of Victoria, 
 3800 Finnerty Rd, Victoria,B.C., V8P 5C2, Canada\\
$^6$ Australia National Telescope Facility, P.O. Box 76, Epping, NSW 2121,
Australia }

\date{Submitted 2009 Oct 19; Accepted 2009 Dec 16}

\abstract{}{Current mass-loss rate estimates imply that main sequence
line-driven winds are not sufficient to strip away the H-rich envelope
to yield Wolf-Rayet (WR) stars. The rich transitional population of
the young massive cluster Westerlund 1 (Wd~1) provides an ideal
laboratory to observe and constrain mass-loss processes throughout the
transitional phase of stellar evolution.}{We present an analysis of
deep radio continuum observations of Wd~1 obtained with the Australia
Telescope Compact Array at four frequency bands that permit
investigation of the intrinsic characteristics of the radio
emission.}{We detect 18 cluster members, a sample dominated by the
cool hypergiants, with additional detections amongst the hotter OB
supergiants and WR stars. The radio properties of the sample are
diverse, with thermal, non-thermal and composite thermal/non-thermal
sources present. Mass-loss rates determined for stars with partially
optically thick stellar winds are
$\sim10^{-5}$\,M$_{\odot}$\,yr$^{-1}$ across all spectral types,
insufficient to enable the formation of WRs during a massive star
lifetime, and the stars must undergo a period of greatly enhanced mass
loss. The sgB[e] star W9, the brightest radio source in Wd~1, may
provide an example, with a current mass-loss rate an order of
magnitude higher than the other cluster members, and an extended
nebula interpreted as a wind from an earlier epoch with a density
$\sim3\times$ the current wind. Such an envelope structure in W9 is
reminiscent of luminous blue variables, and one that shows evidence of
two eras of high, possibly eruptive mass loss. Surprisingly, three of
the OB supergiants are detected, implying unusually dense winds,
though they are embedded in more extended emission regions that may
influence the derived parameters. They also may have composite
spectra, suggesting binarity, which can lead to a higher flux than
expected from a stellar wind.  Spatially resolved nebulae are
associated with three of the four RSGs and three of the six YHGs in
the cluster, which are due to quiescent mass loss rather than
outbursts.  The extended nebulae of W20 and W26 have a cometary
morphology, implying significant interaction with either the
intracluster medium or cluster wind. For some of the cool star winds,
the ionizing source may be a companion star though the cluster
radiation density is sufficiently high to provide the necessary
ionizing radiation. Five WR stars are detected with composite spectra,
interpreted as arising in colliding-wind binaries.}{}

\keywords{stars:evolution - ISM:H\,{\sc ii} regions - Galaxy:Open clusters and 
associations:individual: Westerlund 1} 

\offprints{sean.dougherty@nrc.ca}

\maketitle

\authorrunning{S.M.~Dougherty et al.}  


\section{Introduction}
\label{sec:intro}
Recent work strongly suggests that canonical mass-loss rates for O
stars need to be revised downwards to accommodate the effect of wind
clumping \citep{Fullerton:2006, Mokiem:2007}. Such main sequence (MS)
mass-loss rates are insufficient to remove the H-rich mantle of the
star prior to it becoming a Wolf Rayet (WR), shifting the burden of
mass loss onto the short lived transitional phase of stellar
evolution. This phase is populated by a wide variety of highly
luminous, hot supergiant B[e] and luminous blue variable (LBV) stars,
and cool Yellow Hypergiant (YHG) and red supergiant (RSG)
stars. However the exact path of an O star through this transitional
`zoo' as a function of initial mass is currently poorly understood,
while the short-lived epochs, and hence rarity, of such stars
complicates efforts to constrain their properties such as mass-loss
rate and lifetime.

A better understanding of such short-lived phases in the life cycle of
massive stars is crucial to areas of astrophysics other than just
stellar evolution. For example massive stars are thought to
predominantly form in stellar aggregates, where they drive cluster
winds, which are a major source of mechanical feedback and chemically
processed material into the wider galactic environment, in turn
driving star formation and galactic evolution. Indeed
the feedback from large populations of Super Star Clusters (SSC;
M$_{\rm total} \sim 10^5-10^7$M$_{\odot}$) may drive galactic scale
outflows, which, if present in Dwarf galaxies, may be sufficient to
strip them of their interstellar medium, preventing subsequent
generations of star formation \citep[e.g.][]{Westmoquette:2007}.

One such cluster in the Galaxy is \object{Westerlund 1}
\citep{Westerlund:1961}, hereafter Wd~1, for which photometric and
spectroscopic observations suggested a unique population of both cool
and hot supergiants \citep{Borgman:1970, Westerlund:1987}. Recently,
detailed optical and near-IR observations have confirmed these results
and revealed Wd~1 to be even more extreme than previously anticipated,
containing a large population of post-MS stars with representative
members of all evolutionary stages: OB supergiants and hypergiants,
RSGs, YHGs and WRs \citep[][henceforth C02 and C05
respectively]{Clark:2002, Clark:2005}. Indeed, Wd~1 contains 6 YHGs,
more than 50\% of the currently known population in the Galaxy, as
well as one of the largest WR populations of any cluster in the Galaxy
\citep{Crowther:2006}. With a cluster mass of $\sim$10$^5$M$_{\odot}$
(C05), Wd~1 is directly comparable to the SSCs observed in external
galaxies such as M82 and thus represents a relatively nearby example
that provides a valuable opportunity to study the properties,
evolution and interaction of massive stars in their `natural'
environment.

Radio continuum observations are a long established tool for
estimating mass-loss rates for early-type stars. As part of a
programme to accomplish this for classical Be and B[e] stars,
\cite[][hereafter C98]{Clark:1998} imaged Wd~1 at radio wavelengths
and found two unusually radio luminous stars; the sgB[e] star
\object{W9} and the RSG \object{W26}. In both cases the emission was
found to be spatially resolved, suggestive of recent mass-loss events.
Motivated by these unusual radio properties and the possibility of
detecting emission from massive post-MS stars over a broad range of
evolutionary stages we carried out a more extensive radio observation
of Wd~1 at four frequencies.

We present the results of this survey in this paper. The
observations are described in section~\ref{sec:observations}, with the
radio sources identified in section~\ref{sec:radio_sources}.
Section~\ref{sec:stellarsources} discusses the nature of the radio sources,
section~\ref{sec:diffuse} is a brief discussion of the extended emission,
and the results are summarized in section~\ref{sec:summary}.

\section{Observations}
\label{sec:observations}

Observation of \object{Wd~1} have been obtained at 8640, 4800, 2496
and 1384 MHz (corresponding to 3, 6, 13 and 20cm respectively) using
the Australia Telescope Compact Array. A first epoch was obtained 1998
March 4-5 in the 6B configuration, followed by observations 2001
January 7-8 in the more compact 750C configuration and on 2002 May 18
(only 8640 and 4800 MHz) in the 6A configuration.  Together, these
observations ensure good coverage across the spatial frequency range
observed, though do not have identical spatial frequency coverage at
each frequency. Hence, caution is suggested when comparing flux
measures at the different frequencies for resolved, extended emission.

\begin{figure*}[t]
\vspace{18.9cm}
\includegraphics{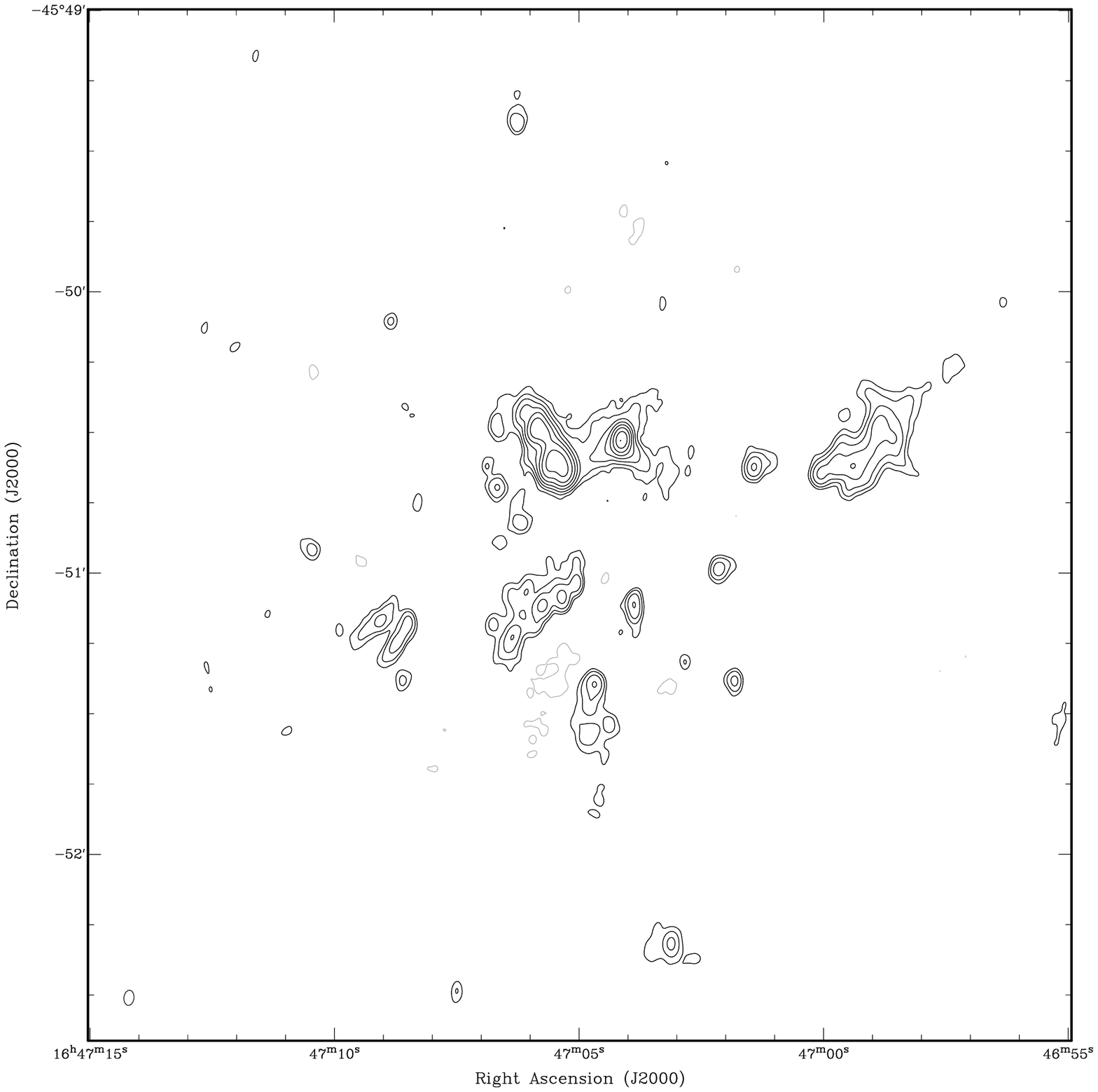}
\includegraphics{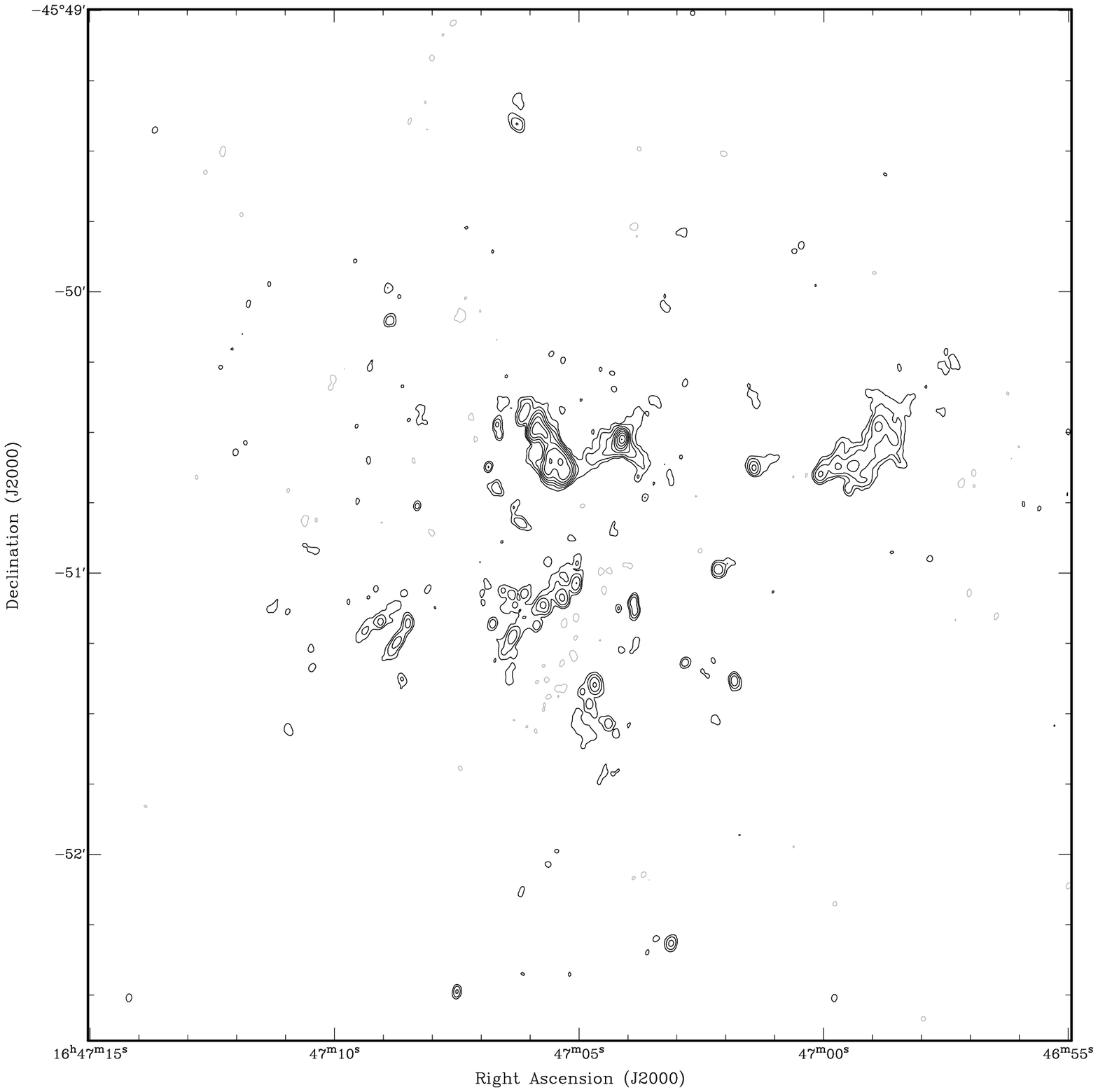}
\includegraphics{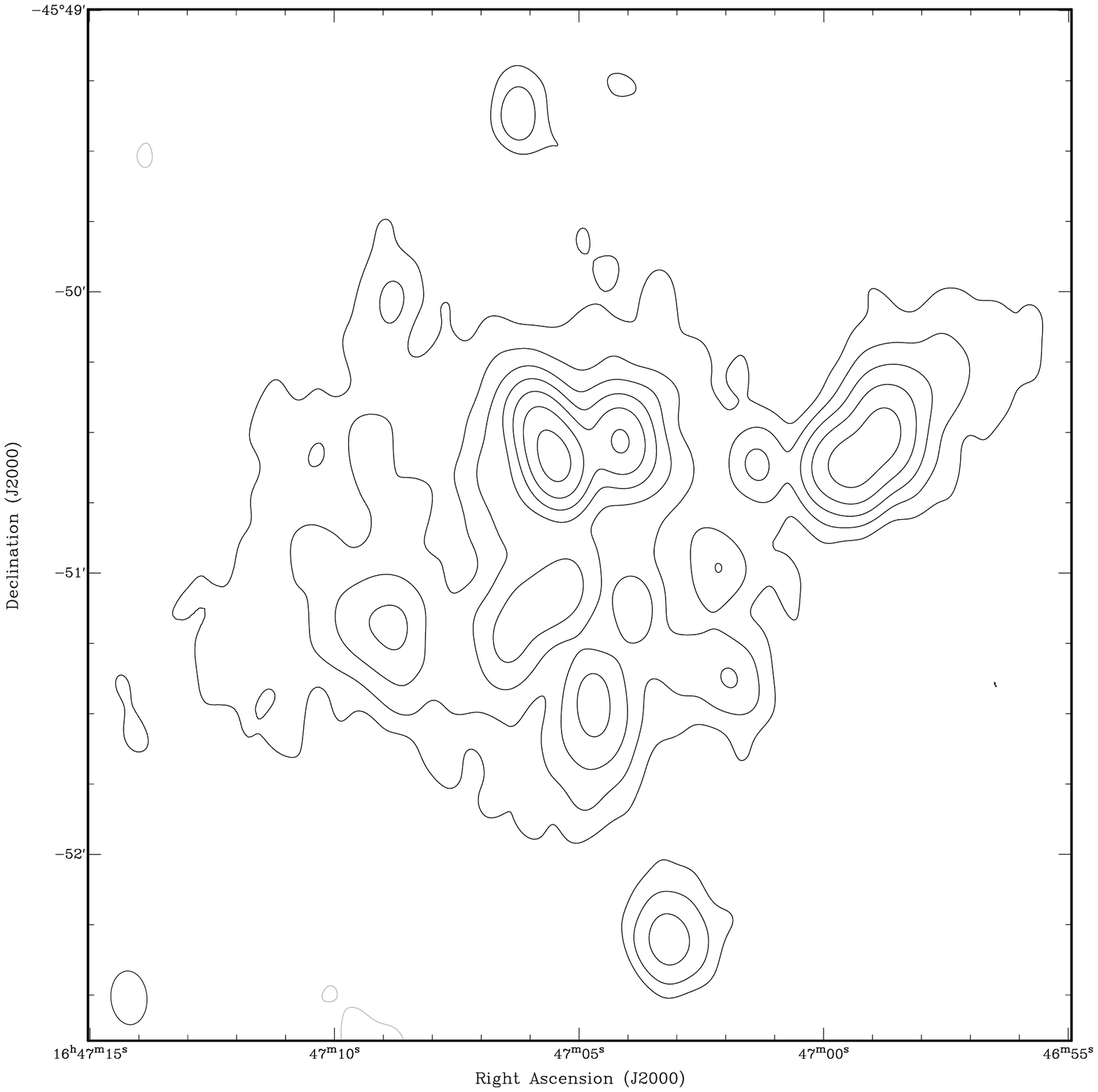}
\includegraphics{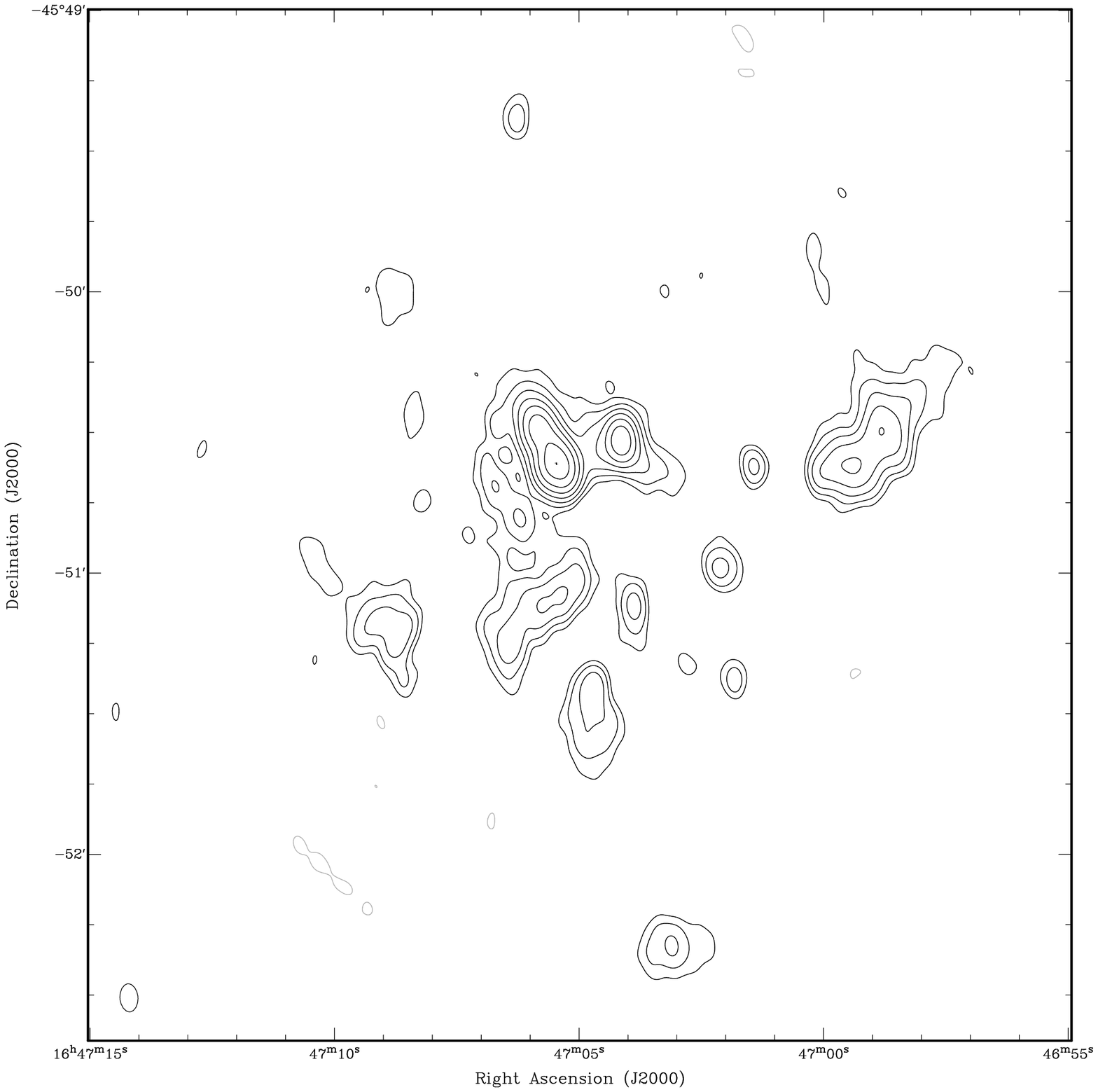}
\caption[] {The ATCA observations of Wd~1 at 8.6~GHz (upper left),
  4.8~GHz (upper right), 2.5~GHz (lower left) and 1.4~GHz (lower
  right). In each image, contour levels are
  -3,3,6,12,24,48,96,$192\sigma$, where the $1\sigma$ rms uncertainty
  is $0.06, 0.1, 0.2, 0.4$~mJy~beam$^{-1}$ respectively at the four
  frequencies. These images were reconstructed using the nominal FWHM
  of the synthesized beam at each frequency, shown in the lower
  left-hand corner of each image. Note that these images are not
  corrected for the primary beam pattern.}
\label{fig:radio_images}
\end{figure*}

Observations of \object{Wd~1} were interleaved with repeated
observations of the nearby bright point source \object{1657-56} in
order to establish the antenna-based complex gains. Initial editing
and calibration of all observations was done using the {\sc Miriad}
data reduction package~\citep{Sault:1995}. The gain solutions
established using \object{1657-56} were subsequently ``referenced'' by
interpolation to the \object{Wd~1} observations.  The absolute flux
scale was established using the primary calibrator \object{1934-638},
assuming fluxes of 2.84, 5.83, 11.14 and 14.94 Jy at 8.6, 4.8, 2.5 and
1.4~GHz respectively. The fluxes derived for \object{1657-56} are
shown in Table~\ref{tab:1657}. No significant flux variations
($>10\%$ of the source flux) were noted in preliminary analysis of the
fluxes of the brightest unresolved sources ($>2$~mJy at 8.6~GHz)
in data from each epoch. Hence, data from the three epochs were
combined into one dataset in order to improve the signal-to-noise
ratio of the data and improve our ability to detect weaker sources.

After phase-referencing and subsequent combination of the data from
each epoch, deconvolution was done through visibility model fitting, a
technique widely used in VLBI image construction, rather than the
commonly used {\sc clean} technique. This was done using an automated
modelling fitting routine, the {\sc modcons} macro within the {\sc
smerf} patch \citep{Reid:2006} to the {\sc Difmap} package
~\citep{Shepherd:1997}. The major advantage of the ``smear fitting''
method implemented by {\sc smerf} is that it yields higher resolution
for {\it significantly} detected features than estimated by the
oft-used Rayleigh criteria \citep{Reid:2006}, represented by the FWHM
of the synthesized beam. A model was established for the 8.6-GHz
visibilities consisting of point and Gaussian source models, using the
image data to guide the modelling process. This gave a model at the
highest observed frequency, and hence at the highest resolution. Once
the positions of the model components had been determined, the
relatively high signal-to-noise ratio of the brightest sources in the
field permitted the use of phase-only self-calibration to improve the
antenna gain solutions from those derived from phase referencing
alone. A ``best'' model was established with a final round of model
fitting, keeping the positions fixed, to refine the flux of the model
components. Note that no amplitude self-calibration was carried out
using these models. This ``best'' model at 8.6~GHz was then used as
the input model at the other frequencies. The resulting images at the
four passbands are shown in Fig.\ref{fig:radio_images}.

\begin{table}
\caption[]{Fluxes determined for the phase-reference 1657-56}
\begin{tabular}{llllll}
\hline
& Config.& 8.6~GHz & 4.8~GHz & 2.5~GHz & 1.4~GHz \\
& &(Jy)& (Jy)& (Jy)& (Jy)\\

\hline
1998 Mar & 6B & $1.54$ & $1.81$ & $1.92$ & $1.89$ \\
2001 Jun & 750C & $1.44$ & $1.61$ & $1.71$ & $1.88$ \\
2002 May & 6A & $1.63$ & $1.79$ & &\\
\hline
\end{tabular}
\label{tab:1657}
\end{table}

\begin{figure*}
\vspace{17.3cm}
\includegraphics{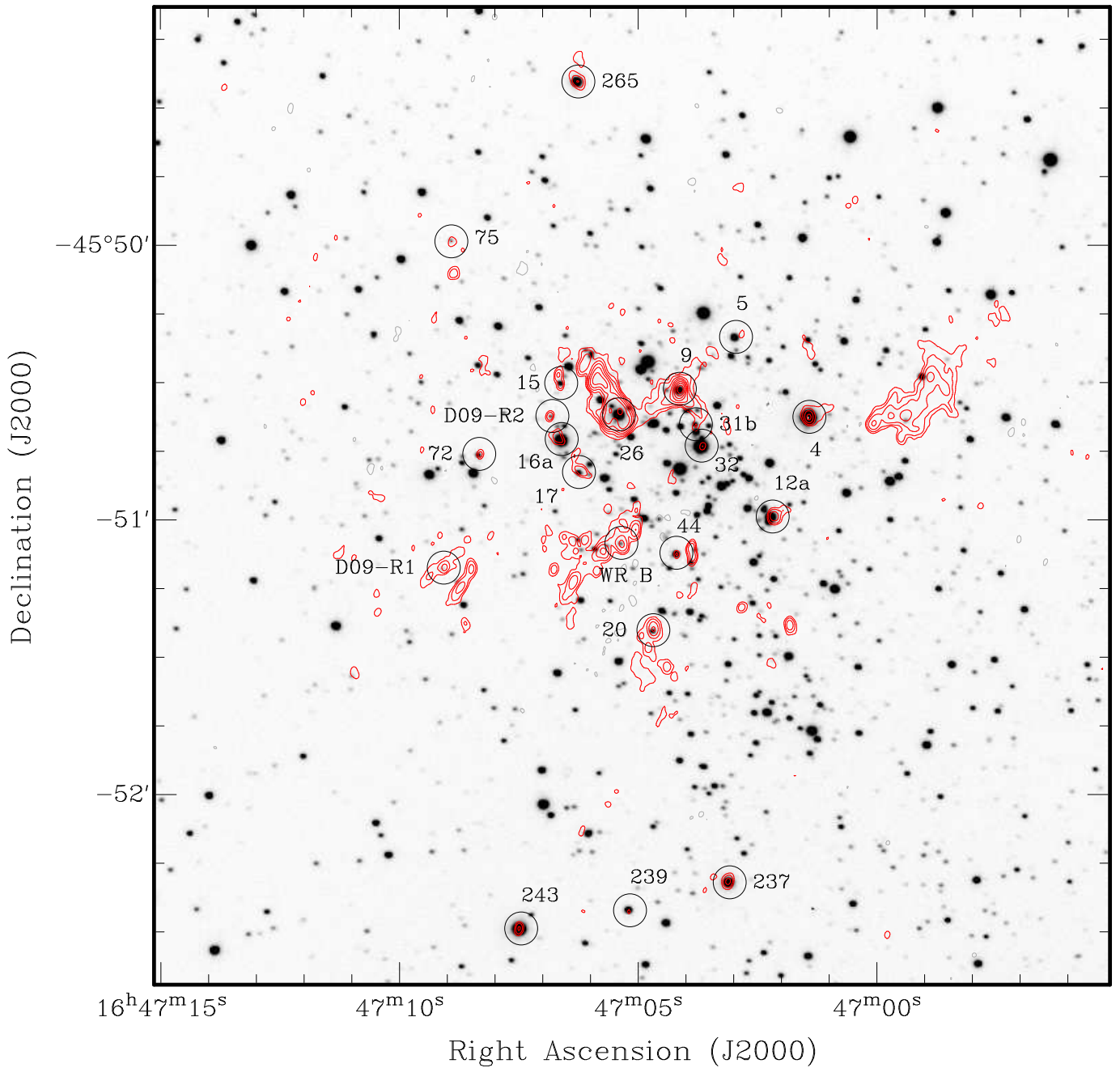}
\caption[] {8.6-GHz image overlaid on a FORS $R$-band image. The
limiting magnitude of the $R$-band image is
$\approx17.5$~magnitudes. The radio data are represented by contours
with levels at -3,3,6,12,24,48,96,$192\sigma$, where the $1\sigma$ rms
uncertainty is $0.06$~mJy~beam$^{-1}$. The radio sources with putative
optical counterparts listed in Table~\ref{tab:radiometry} are
identified by the circles and Westerlund numbers.}
\label{fig:fors2_3cm}
\end{figure*}

\begin{table*}
\caption[]{Characteristics of stellar radio sources in the \object{Wd~1} cluster}
\label{tab:radiometry}
\begin{tabular}{|l|c|l|l|l|l|l|ll|l|c|}
\hline
Source ID & Spectral & 8.6~GHz$^d$ & 4.8~GHz  & 2.5~GHz  & 1.4~GHz &  Spec. Index &
\multicolumn{2}{c|}{Radio pos.$^e$}&$\Delta^f$&Notes\\
$^a$$^{(b)}$& Type$^c$ & (mJy)&(mJy)&(mJy)&(mJy)&&$16^{\rm h}47^{\rm m}$ & 
$-45^{\circ}$&($\prime\prime$)&\\
\hline
 4(H)  &           & $0.8\pm0.08$  & $0.6\pm0.1$ &      &              &$+0.49\pm0.31$ &1.445&50~37.66&0.37&\\
 4&\raisebox{1.5ex}[0pt]{F2 Ia$^+$}&  $2.2\pm0.2$r$^h$  & $2.8\pm0.3$r &\raisebox{1.5ex}[0pt]{$3.3\pm0.3^g$} &\raisebox{1.5ex}[0pt]{$3.8\pm0.4$} &$-0.23\pm0.07$&&&&1\\
 9(C)  &           & $24.9\pm2.5$ & $18.1\pm1.8$ & $9.5\pm0.9$ & $7.8\pm0.8$  &$+0.68\pm0.07$&4.136&50~31.41&&\\
 9&\raisebox{1.5ex}[0pt]{sgB[e]}& $30.5\pm3.0$r& $28.4\pm2.8$r&$21.0\pm2.1$r&$24.5\pm2.5$r&$+0.16\pm0.07$&&&&\\
12a    & A5 Ia$^+$ &  $2.9\pm0.3$r & $2.9\pm0.3$ &  $3.5\pm0.4$ & $3.0\pm0.4$ &$-0.06\pm0.08$&2.151&50~59.23&0.61&1\\
15     & O9.5-B0.5Ia     &  $0.6\pm0.06$  & $0.7\pm0.1$ &$<0.6^i$&$<1.1$      &$>0.0$&6.648&50~30.31&0.54& \\
17     & O9.5-B0.5Ia     &  $1.7\pm0.2$  & $1.6\pm0.2$ &$3.2\pm0.3$ & $2.9\pm0.4$   &$-0.42\pm0.11$&6.183&50~49.40&0.50&\\
WR B& WN7     &  $4.3\pm0.4$r  & $4.8\pm0.5$r&$4.4\pm0.4$ & $4.0\pm0.4$   &$+0.04\pm0.07$&5.345&51~\phantom{1}5.21 &0.21&2\\
20     & RSG       &  $3.8\pm0.4$r & $4.2\pm0.4$r&$4.1\pm0.4$& $4.8\pm0.4$    &$-0.11\pm0.07$&4.679&51~23.81&0.33&2\\
26(A)  & RSG       & $20.1\pm2.0$r & $21.9\pm2.2$r& $18.4\pm1.8$r& ?          &$+0.07\pm0.11$&5.370&50~36.57&0.33&2\\
31b(WR V)& WN8      & $ 0.4\pm0.06$  & $0.4\pm0.1$&$<0.6$&$<1.1$          &$\phantom{+}0.00\pm0.50$&3.821&50~39.22&0.45&\\
32     & F5 Ia$^+$ & $0.4\pm0.06$  &  $0.4\pm0.1$ &$<0.6$&$<1.1$        &$\phantom{+}0.00\pm0.50$&3.675&50~43.70&0.31&\\
44(WR L) & WN9     & $0.4\pm0.06$  &$<0.3$&$<0.6$&$<1.1$        &$>0.5$&4.210&51~\phantom{1}7.34 &0.23&\\
D09-R1   &           & $0.7\pm0.07$  & $0.8\pm0.1$ &      &              &$-0.23\pm0.27$&9.075&51~10.23&0.31&\\
D09-R1? &\raisebox{1.5ex}[0pt]{BSG}&$6.5\pm1.2$r&$6.2\pm1.2$r&\raisebox{1.5ex}[0pt]{$11.0\pm1.2$r}&\raisebox{1.5ex}[0pt]{$17.2\pm1.7$r} &$-0.61\pm0.11$&&&&1,3 \\
72(WR A) & WN7   &  $0.5\pm0.06$  &  $0.5\pm0.1$ &$<0.6$&$<1.1$        &$\phantom{+}0.00\pm0.40$&8.332&50~45.58&0.00&\\
75     &  RSG      &  $0.3\pm0.06$  &$<0.3$&$<0.6$&$<1.1$        &$>0.0$&8.933&49~58.92&0.46&\\
237(B) &           &  $1.8\pm0.2$  &  $1.8\pm0.2$ &  $1.9\pm0.2$ &  $2.0\pm0.4$         &$-0.06\pm0.10$&3.117&52~18.99&0.28&\\
237    &\raisebox{1.5ex}[0pt]{RSG}&  $5.6\pm2.2$r&  $7.5\pm2.1$r& $10.6\pm3.0$r& $11.2\pm1.9$r&$-0.35\pm0.19$&&&&1,3\\
239(WR F) &  WC9d  &  $0.3\pm0.06$  &$<0.3$&$<0.6$&$<1.1$        &$>0.0$&5.196&52~25.63&0.50&\\
243(G) & LBV       &  $1.5\pm0.2$  &$0.9\pm0.1$ &$<0.6$&$<1.1$        &$+0.87\pm0.30$&7.496&52~29.28&0.32&\\
265(I) & F5 Ia$^+$ &  $2.3\pm0.3$r &$ 2.1\pm0.2$ & $1.7\pm0.2$ &  $1.8\pm0.4$   &$+0.20\pm0.11$&6.268&49~24.19&0.30& \\
\hline
 5(WR S)& WN10-11 & $0.3\pm0.06$   &$<0.3$&$<0.6$&$<1.1$        &$>0.0$&2.877&50~19.76 &0.91&4\\
16a(L) & A2 Ia$^+$ &  $1.6\pm0.3$ & $2.0\pm0.2$ &$2.8\pm0.3$ & $3.0\pm0.4$   &$-0.33\pm0.09$&6.688&50~41.99&0.93&4\\
D09-R2   & BSG   &  $0.7\pm0.06$  & $0.8\pm0.1$ &$1.2\pm0.2$ & $1.6\pm0.4$   &$-0.43\pm0.11$&6.867&50~37.17&0.70&4\\
\hline
\end{tabular}

a) Source numbers from \cite{Westerlund:1987}. \\ 
b) Alternate names from C02, \cite{Borgman:1970},\cite{Negueruela:2005}, and \cite{Crowther:2006}\\
c) Spectral types from C02, C05, \cite{Negueruela:2005} and \cite{Crowther:2006}\\ 
d) Fluxes derived from visibility modelling, corrected for the primary beam of the ATCA.\\ 
e) J2000 positions of model components at 8.6~GHz, assuming reference source 1657-56 at $\alpha$ = 17:01:44.853, $\delta$ =
-56:21:55.964. The uncertainty in absolute radio position is estimated to be $\pm0.1$ arcsec\\ 
f) Separation of the optical and radio positions. \\
g) For sources with two components at higher frequencies, when there is only one model component determined this 
represents a total flux at this frequency \\
h) The symbol ``r'' denotes components that appear resolved by the ATCA at that frequency.\\ 
i) Quoted upper limits are $3\sigma$.\\ 
1) Extended component \\
2) These sources are embedded in extended emission which may influence the flux estimates. \\
3) Not clear if the extended emission is related to the compact radio source or the underlying optical source.\\
4) The radio emission is offset from the stellar position by more than $3\times$ the position uncertainty introduced by phase-transfer i.e. 0.6 arcsec. Hence, association of radio emission with the optical source is uncertain.\\ 

\end{table*}

\section{The radio sources in \object{Wd~1}}
\label{sec:radio_sources}
\subsection{Identification of radio stars}
In order to identify stellar counterparts to the radio sources in
\object{Wd~1}, an $R$-band image (655 nm effective wavelength) of the
cluster was obtained on 2004 June 13 using the FORS2 camera on the
VLT.  After re-projecting the optical image to the same projection as
the ATCA images using the {\sc aips} task {\sc ohgeo}, the absolute
position of the optical image was fixed by assuming that the position
of the point source component of the 8.6-GHz emission from
\object{W9}, the brightest source in the radio image, is coincident
with the peak of its optical emission, as determined using {\sc aips}
routine {\sc maxfit}.  Since the position of 8.6-GHz emission
components was established by phase-referencing, the absolute position
accuracy of the model components is determined by the accuracy of the
position of the phase-reference source 1657-56, and on the residual
antenna gain phase as a result of the offset between the position of
the phase-reference source and Wd~1, separated by 8.8 degrees. It is
estimated this leads to an uncertainty of $\sim100$~mas in the
absolute positions of the radio sources (see
Table~\ref{tab:radiometry}).

The resulting overlay of the 8.6-GHz image and the $R$-band image is
shown in Fig~\ref{fig:fors2_3cm}.  A large number of optical sources
are coincident with radio sources. Setting a $4\sigma$ point source
detection limit of $0.23, 0.43, 0.86$ and $1.45$~mJy~beam$^{-1}$ at
8.6, 4.8, 2.5 and 1.4~GHz respectively, we identify the optical
sources with associated radio emission in Table~\ref{tab:radiometry}.

The position of putative optical counterparts to the radio sources were
also determined using the {\sc aips} routine {\sc maxfit}, and an
offset between the radio and optical positions derived. The
uncertainty in the relative position of the radio and the optical
image is governed by how well the optical image is referenced to the
radio image. By inspection of the position of \object{W9}, we estimate
this to be an uncertainty of approximately $\pm200$~mas. This is the
dominant position uncertainty for the sources. Radio sources that have
a position offset from a putative optical ``counterpart'' greater than
600~mas may not be associated, and so these potential
mis-identifications are identified in Table~\ref{tab:radiometry}.

A total of 18 stars are identified as radio emitters, with
optical-radio positional offsets $<0.6''$. We identify two objects
with associated or potentially associated radio emission for which no
designation is available in the literature. Spectroscopy that could
identify if they are bona-fide cluster members is not available for
these objects so we do not assign Westerlund numbers to them, but name
them as D09-R1 and D09-R2 following the convention used by
\cite[][hereafter C08]{Clark:2008}. Three stars are identified as possible radio
emitters (W5, W16a and D09-R2) since they are offset from the
potential radio ``counterpart'' by between 0.6 and 1.0 arcseconds, and
some caution may be warranted identifying these stars with the nearby
radio emission.

With this many stellar radio sources, \object{Wd~1} has the richest
population of radio emitting stars known for any young massive
galactic cluster \citep[e.g.][]{Lang:2003, Moffat:2002,
SetiaGunawan:2003}. The stellar radio sources are blue, yellow or red
super- or hypergiants and WR stars, representative of different stages
of massive star evolution.  The supergiant B[e] star W9 is by far the
brightest stellar radio emitter in the cluster, as anticipated from
the ``snapshot'' observation of the cluster presented by
C98. Among the other four known blue hypergiants, only
the LBV W243 is detected.  Wd~1 has one of the richest populations of
YHGs known in the Milky Way, with six members of this group (W4, 8a,
12a, 16a, 32, and 265), all of which are detected except for W8a and
possibly W16a. All four known RSGs, six of the 16 WR stars in the
field-of-view (24 known currently in the cluster), and four OB
supergiants, are also associated with radio emission.

A number of sources appear to consist of multiple components (W4, W9,
D09-R1 and W237) with a central unresolved source surrounded by a more
extended component that is often significantly larger than the
synthesized beam of the array. For W4 and W9, the extended component
is clearly centred on the stellar source and an association with the
extended emission is strongly implied. For D09-R1, the relationship of
the stellar source and the extended emission is not as clear since
the extended emission is offset from the associated point source. 

\subsection{Radio star characteristics}

The radio fluxes of the sources were deduced from the model components
derived in the model fitting process, which were then corrected for
attenuation by the primary beam of the ATCA, dependent on their
distance from the pointing centre. The FWHM of the ATCA primary beam
is $5.86, 10.06, 19.91$ and $34.61$ arcminutes at 8.6, 4.8, 2.5 and
1.4 GHz respectively. Given the size of the field-of-view is
$\sim4$~arcminutes, this leads to flux corrections of up to
$1.4\times$ in the outer regions of the field-of-view at 8.6 GHz.  In
an attempt to allow for systematic errors, the flux uncertainties
quoted in Table~\ref{tab:radiometry} are taken from the maximum of the
following three values: the $1\sigma$ rms uncertainty in the image,
10\% of the source flux, or for resolved sources the rms uncertainty
in the image multiplied by the solid angle of the source. This
represents a conservative approach to estimating the flux
uncertainties, and hence to the estimated spectral indices. However,
it presumes that the model components used to describe a source are
not influenced by other emission from the region surrounding the
source. This may be an additional complication in determining the
fluxes of several sources (WR~B, W20 and W26) since they are embedded in
more extended regions of emission that may impact the accuracy of the
derived fluxes. In addition to the source flux, it is noted which sources
appear resolved from the visibility modelling directly, avoiding 
the impact of visibility weighting during imaging on the {\sc clean}
beam size and the subsequent limitations of source modelling in the
image plane.

It is noted that the flux values determined in this study are
approximately 10\% higher than the values previously reported by
C98. It is not clear if the apparent increase in the
flux of W9 is due to intrinsic source variations or due to the
absolute calibration scale, given that the uncertainty in the absolute
flux calibration is $\sim5-10$\%. We suspect the apparent variation is
due to a combination of absolute calibration uncertainty and the fact
the data used in this study extend to lower spatial frequencies than
in the study of C98, and hence may recover flux that was
``resolved out'' in the earlier observations. We suggest a more
accurate determination of total flux is presented here than in the
previous work.

\begin{figure}
\vspace{6.1cm}
\includegraphics{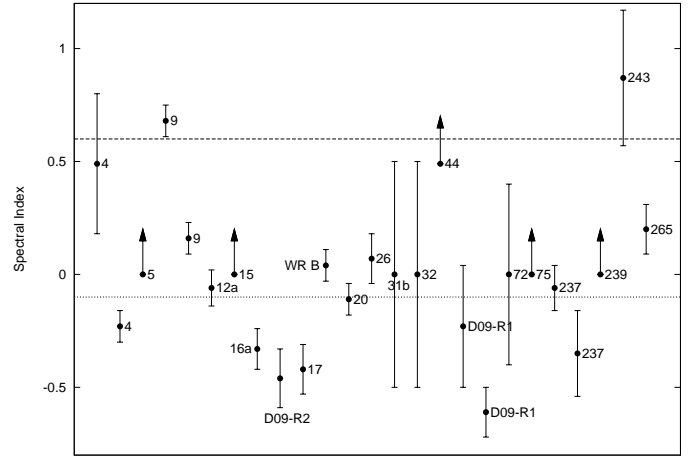}
\caption[] {The spectral indices of the stellar radio sources in Wd~1,
identified by their Westerlund numbers.  The canonical spectral
indices for partially optically thick thermal emission from a stellar
wind ($+0.6$) and optically thin thermal emission ($-0.1$) are
highlighted by the dashed and dotted lines.  Error bars are 1
sigma. The arrows denote lower limits to the derived spectral
index. Radio sources offset from potential optical counterparts by
more than 0.6'' are marked by open boxes. }
\label{fig:spec_index}
\end{figure}

The radio spectral index $\alpha$, where $S_\nu \propto \nu^\alpha$,
for each source was calculated by a weighted regression fit of a
single power-law to the fluxes. Use of a single power-law presumes
there is no curvature in the continuum spectra over the frequency
range observed. Inspection of the spectra of those sources detected at
all bands shows no compelling evidence of curvature, given the
uncertainties in the radiometry at each band. The resulting values for
spectral indices and uncertainties are given in
Table~\ref{tab:radiometry} and displayed in
Fig.~\ref{fig:spec_index}. For sources with radiometry at all
frequency bands, the resulting single power-law index values are less
affected by potential systematic errors in the flux measures at each
band, compared to indices derived across a smaller frequency range
e.g. 8.6 to 4.8 GHz.  For sources that are undetected at one or more
bands, and upper limits quoted, only the detected fluxes were used to
calculate spectral index, unless they helped further constrain the
spectral index value e.g. W44. It should be noted that the fluxes for
sources only detected at one or two bands are all low ($<1$~mJy) and
the uncertainty in the derived spectral indices is high.

None of the sources have a completely optically-thick thermal spectrum
($\alpha\sim2$), though several have spectra consistent with
$\alpha\sim+0.6$, expected for a partially optically-thick,
steady-state stellar wind e.g. the compact components in W4, 9, 44 and
243.

The majority of the sources have indices that are quite shallow
compared to the expected value for a stellar wind, with many having
indices consistent with $-0.1$, the value for optically-thin thermal
emission. A combination of optically thick and thin emission
components can give rise to such spectra, such as clumpy stellar winds
where the clumps are optically thick
\citep[e.g.][]{Ignace:2004}. Alternatively, in some of these objects
the emission may be due to a combination of thermal and non-thermal
emission forming a composite spectrum, as observed in a number of
massive stars, where the non-thermal emission is attributed to a
colliding-wind binary (CWB) \citep[e.g.][]{Chapman:1999,
Dougherty:2000,vanloo:2006}.  Depending on the relative strength of
the two continuum components, the spectral index could lie anywhere
between $\sim+0.8$ and $-0.6$, this latter value typical of
optically-thin synchrotron emission. This makes an unequivocal
identification of the underlying emission difficult. To compound
matters, composite spectra due to thermal$+$non-thermal emission may
not be well represented by a single power-law
\citep[e.g.][]{Pittard:2006}. However, identification of
CWBs is helped through observations at other
wavelengths, largely thermal X-ray emission arising from shock-heated
plasma in the wind-collision region, and/or dust emission that appears
to be a common feature in carbon-type WR star massive binaries with
non-thermal emission \citep[][]{Williams:1999}.

There are a few sources with indices that, at face value, are more
negative than $-0.1$, suggesting non-thermal emission e.g. W17 and the
extended component associated with W4. Other potential members of this
group are W16a, D09-R2 and the extended emission near D09-R1 but, as
noted earlier, the association of the radio emission with the
underlying stars is unclear.  The derived spectral index for the
extended emission in W4 is consistent with optically- thin thermal
emission at the 95\% confidence level. This leaves W17 as the only
bona-fida non-thermal source.

Additional observations over a broader range of frequencies are
required to establish more firmly the nature of the radio emission
from the stars in Wd~1. Individual cases are discussed in
Sec.~\ref{sec:stellarsources} where the observations from each stellar
sub-type are discussed in detail.

\begin{table}
\caption{Number of radio emitters of given spectral type}
\begin{tabular}{lccc}
\hline
Spec.  & No. radio & Source &Cluster \\
Type   & emitters  & ID     &total \\
\hline
OB SGs & 3(4?)$^a$ & 15, 17, D09-R1, (D09-R2?) & $\sim$150  \\
sgB[e] & 1 & 9 & 1 \\
BHGs   & 1 & 243 & 4 \\
YHGs   & 4(5?)$^a$ & 4, 12a, 32, 265 (16a?) & 6 \\
RSGs   & 4 & 20, 26, 75, 237 & 4 \\
WN9-10 & 1 (2?)$^a$ & 44, (5) & 2 \\
WN5-8  & 3 & WR~B, 31b, 72 &14 \\
WC     & 1 & 239 & 8 \\
\hline
\end{tabular}

Notes: a) These sources are included in parentheses due to the
uncertain association of radio emission with W16a, D09-R2 and W5. b) WRs N
(WC8), O and Q (both WNL) are not in the field-of-view of the radio
observations described here.
\end{table}

\begin{table}
\caption[]{Properties of the extended radio emission regions in the
cluster centre.\label{tab:unassociated}}
\begin{tabular}{llllll}
\hline
ID & 8.6 GHz & 4.8 GHz &2.5 GHz & Spec. Index & Assoc.\\
&(mJy)&(mJy)&(mJy)&&Star\\
\hline
A1 &  5.6 &  7.6 &  7.6 &$-0.24\pm0.15$  & D09-R1\\
A2 & 23.8 & 26.0 & 35.5 &$-0.32\pm0.10$  & WR~B\\
A3 & 12.1 & 11.1 & 16.0 &$-0.23\pm0.20$  & W20 \\
A4 &  3.5 &  4.7 &  5.5 &$-0.36\pm0.10$  & \\
A5 &  2.0 &  1.9 &  2.0 &$-0.01\pm0.11$  & \\
A6 & 71.9 & 78.0 & 81.8 &$-0.05\pm0.07$  & \\
A7 & 22.4 & 28.2 & 27.0 &$-0.14\pm0.13$  & W9\\
A8 &153.3 &164.9 &161.4 &$-0.04\pm0.11$  & W26\\
Total&295&322&337&&\\
\hline
\end{tabular}
Note: These fluxes are for the extended emission only in these
regions. The fluxes of the radio stars in these regions has been
subtracted. The flux uncertainty is assumed to be $\sim10\%$.
\end{table}

\subsection{Extended radio emission}
\label{sec:extended}
\begin{figure}[t]
\vspace{8.5cm}
\includegraphics{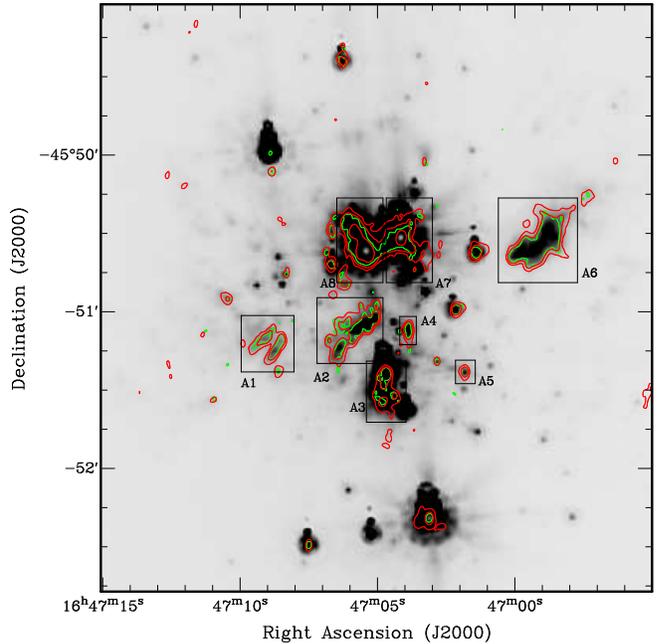}
\caption[] {Comparison of radio emission at 8.6 (green) and 4.8-GHz
(red) observations with $8\mu$m emission taken as part of the Spitzer GLIMPSE
project. Only a few radio contours that demark the extended radio emitting
regions are shown for clarity.}
\label{fig:dustoverlay}
\end{figure}

In addition to the radio emission that is associated with stars it is
clear there are a number of large, extended emission regions,
distributed within $\sim1.5$ arcminutes of the cluster centre. Some of
these regions appear to be associated with stars e.g.  A3 with W20 and
A8 with W26, but there are others such as A5 and A6 for which optical
counterparts are not identified in the FORS2 image. These regions are
identified in Fig.~\ref{fig:dustoverlay} and the radio properties
summarised in Table~\ref{tab:unassociated}.  The spatial frequency
coverage is not the same at all frequencies and some extended emission
may be resolved out at 8.6 GHz and hence the reported fluxes may be
low relative to the fluxes at lower frequencies. Hence, some of the
spectral indices reported for the individual regions in
Table~\ref{tab:unassociated} may actually be flatter, or less
negative, than reported.

The {\em total} radio flux from Wd~1 as measured in the interferometer
data presented here is 422, 461, 523 and 669 mJy at 8.6, 4.8, 2.2, and
1.4 GHz respectively. This is compared to an interpolation of the
single-dish fluxes reported by \cite{Kothes:2007}, which gives fluxes
at the same frequencies of 450, 499, 560, and 620~mJy. Given that the
flux uncertainties in the single dish measures is likely $\sim$10\%,
the total fluxes derived here from the ATCA data are in excellent
agreement with the single dish fluxes.

To determine the flux in extended emission, we subtract the radio
emission associated with the stellar sources as given in
Table~\ref{tab:radiometry} from the total flux determined by the ATCA.
This leads to fluxes of 307, 351 and 426~mJy at 8.6, 4.8 and 2.2~GHz,
with an uncertainty of $\sim5-10\%$ . The flux at 1.4 GHz is not
quoted since it is difficult to ascertain accurately for the radio
stars, most especially for sources embedded in extended
emission. Comparing these values to those given in
Table~\ref{tab:unassociated} and taking into account their
uncertainty, shows good agreement, perhaps degrading toward 2.2 GHz,
which suggests there is little other extended emission in Wd~1 beyond
the regions demarked in Fig.~\ref{fig:dustoverlay}.

Using the total fluxes above for the extended emission in Wd~1 gives a
spectral index of $-0.26\pm0.07$, consistent with optically-thin
thermal emission.
Following the prescription given in Sec.~\ref{sec:cool}, an 8.6-GHz
flux of 310~mJy from a region $1.5\arcmin$ in diameter leads to an
ionized mass of 15~M$_\odot$ for a plasma temperature of 10~kK, in
excellent agreement with the ionized mass derived by
\cite{Kothes:2007}.

C98 showed that the nebular radio emission associated with the RSG W26
correlated with mid-IR emission.  Figure~\ref{fig:dustoverlay} shows
an overlay of the radio emission in Wd~1 with Spitzer GLIMPSE data at
$8\mu$m \citep[][]{Benjamin:2003}. Without being distracted by the
saturation and artefacts in the mid-IR image, it is striking that all
the extended radio emission is associated with mid-IR emission.
Additionally, note that mid-IR emission is associated with each of the
RSG and YHG stars in the cluster (Sec~\ref{sec:cool}). Such a close
correlation between the ionized emission evidenced in the radio and
the 8$\mu$m emission is well established in ultra-compact H{\sc ii}
regions, where the 8$\mu$m emission is due to excited PAH emission
\citep[e.g.][]{Hoare:2007} whereas in more evolved H{\sc ii} regions
there is a clear association with hot dust, and the PAH emission
associated with the photo-dissociation region beyond the ionized gas
\citep[e.g.][]{Povich:2007, Watson:2008}

\section{The stellar sources}
\label{sec:stellarsources}

The post-MS evolution of stars with masses in excess of
$\sim$30M$_{\odot}$ is characterized by significant mass loss. While
such mass loss plays a critical role in determining the subsequent
evolution of the stars, the physical processes driving it are
currently ill-constrained, with the mass-loss rate and wind velocities
anticipated to differ by several orders of magnitude depending on the exact
evolutionary phase.  Given that Wd~1 is uniquely well stocked with
examples of every known post-MS stellar type, these observations offer
the possibility of constraining this process.

Consequently, the following discussion on the nature of the radio
emission for the 18 stellar detections is presented in a likely
evolutionary sequence from OB supergiants, through cool
super-/hypergiants, early hypergiants and WN and WC Wolf-Rayet
stars. Given their comparatively weak winds
($\sim10^{-7}$~M$_\odot$~yr$^{-1}$), the mid-to-late O-type main
sequence progenitors of these stars are not detected
\citep{Martins:2005}.  The emission associated with the sgB[e] star W9
is discussed outside this evolutionary sequence, due to the uncertain
placement of the sgB[e] phase in the post-MS stellar zoo.

Throughout the discussion it will be assumed that Wd~1 is at a
distance of 4~kpc. Recent deep-IR imaging identifies the MS and pre-MS
populations of Wd~1, leading to a photometric distance of
$4.0\pm0.3$~kpc \citep{Brandner:2008}, consistent with both
$4.7\pm1.1$~kpc from an analysis of near-IR photometry of the WR stars
in Wd~1 \citep{Crowther:2006}, and on an atomic hydrogen absorption
distance of $3.9\pm0.7$~kpc \citep{Kothes:2007}.

\subsection{W9 - a luminous radio source}
\label{sec:w9}
By far the brightest radio source in Wd~1 is W9, with a total flux at
8.6~GHz of 55.4 mJy, giving an 8.6-GHz luminosity of
$1.6\times10^{21}$ erg s$^{-1}$. This places W9 among the most radio
luminous thermal emitting stars known: a factor of a few less than the
8.6-GHz luminosity of the extreme LBV $\eta$ Carina
($4.5\times10^{21}$ erg s$^{-1}$) at radio minimum
\citep[$\sim600$~mJy;][]{Duncan:2002}, comparable to the massive
overcontact binary \object{RY Scuti} \citep{Gehrz:1995} and a factor
of a few more luminous than the other oft-cited radio luminous thermal
emitter MWC 349A \citep[e.g.][]{Tafoya:2004}.

C98 hypothesised that W9 comprised of two components, a
compact stellar wind source and an extended region.  The observations
presented in this paper support this model, with an unresolved source
surrounded by an extended region of emission that are both detected at
all four observing frequencies.

The spectral index of the unresolved source coincident with W9 is
$+0.68\pm0.08$, consistent with thermal emission arising from a
partially optically thick, steady-state stellar wind with an $r^{-2}$
radial density distribution. It is assumed this component represents
the present day stellar wind from W9.  The spectral index of the
extended region is $+0.16\pm0.07$, which we take to be essentially
flat, and arguably consistent with optically-thin thermal
emission. Assuming the extended region has a $r^{-2}$ radial ion
density distribution, similar to the stellar wind, the lack of a
turnover in its continuum spectrum to a positive spectral index
implies that the extended region is totally optically thick down
to 1.4~GHz. With the free-free opacity for an $r^{-2}$ ion
distribution at frequency $\nu$ behaving as $\tau_\nu(p)\propto
p^{-3}\nu^{-2.1}$ where $p$ is the impact parameter (see
Appendix~\ref{sec:appendix} and Eqn.~\ref{eqn:opacity}), the lack of a
turnover implies that the extended region must have an inner radius
that is larger than the radius of the $\tau_\nu=1$ surface at 1.4
~GHz. In this case, it is suggested that the extended region
represents an earlier phase of mass-loss from W9, prior to the start
of the current stellar wind phase.

\begin{figure}[t]
\vspace{12.1cm}
\includegraphics{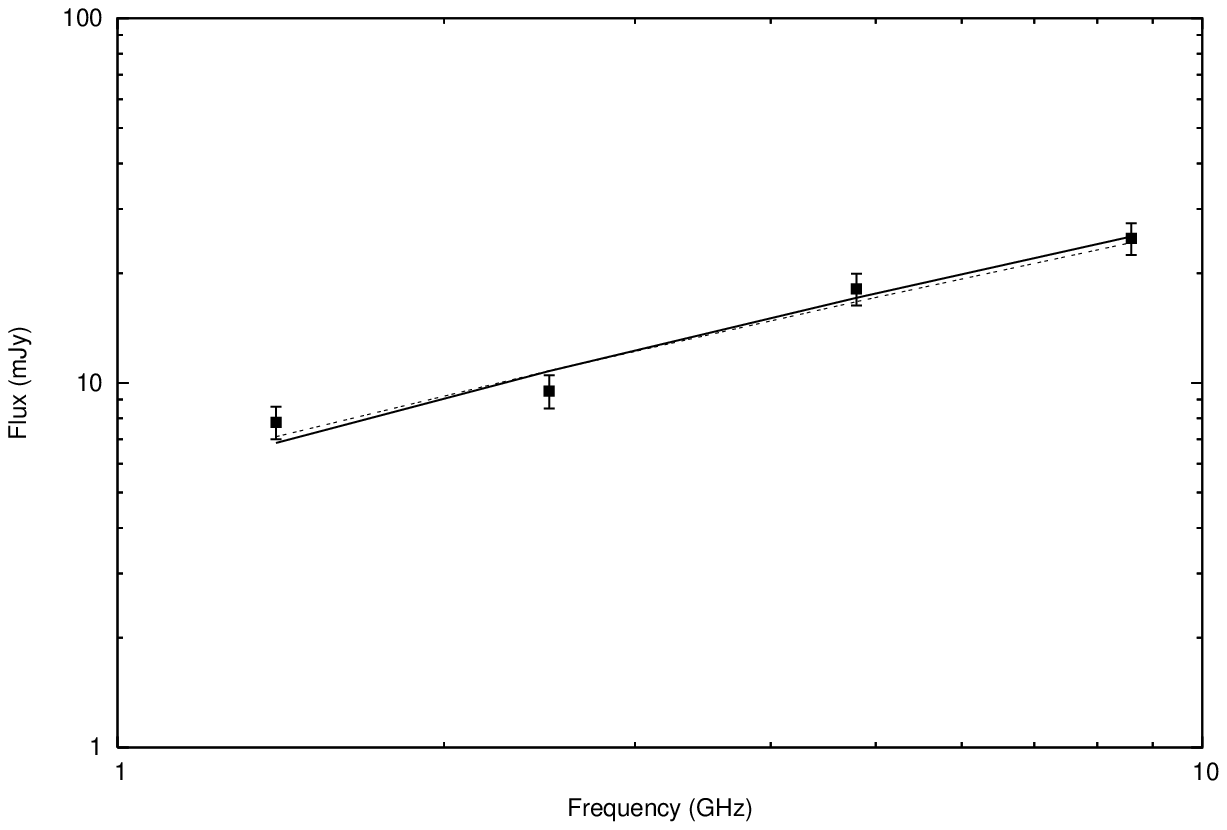}
\includegraphics{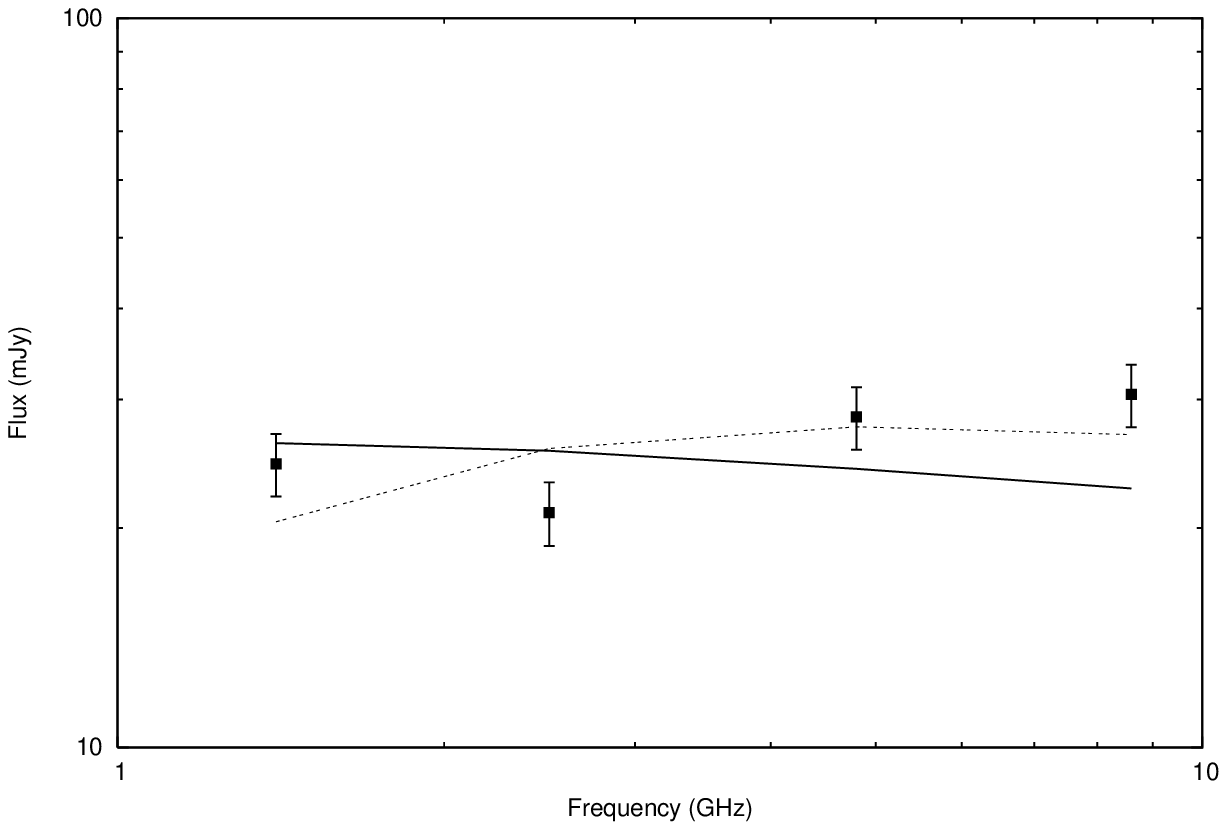}
\caption[]{Model fits to the continuum spectra of the unresolved
partially-thick stellar wind (top) and extended, optically-thin
(bottom) components of W9. The solid lines represent the best-fit
model when the stellar wind outer radius/extended inner radii are
constrained to be identical (Model 3). The dotted
spectra are the best-fit models when the stellar wind outer radius
and extended region inner radius are free parameters (Models 1 and
2).}
\label{fig:w9models}
\end{figure}

To model the continuum spectra of both the stellar wind and the
extended envelope a shell-like geometry is adopted, with outer and
inner radii $R_o$ and $R_i$ respectively and with a $r^{-2}$ radial
ion distribution \citep[e.g.][]{Taylor:1987}. The free-free opacity at
radius $p$ in such an envelope is determined using
Eqns.~\ref{eqn:opacity},~\ref{eqn:opacity_solutions1} and
~\ref{eqn:opacity_solutions2}. Together with Eqn~\ref{eqn:flux}, these
lead to the flux from the envelope at frequency $\nu$. Assuming an
electron temperature $T_e=10$~kK and an outflow velocity of
200~km\,s$^{-1}$, a best-fit of this model to the continuum spectra of
the two components over the four observing frequencies gives the
parameters of the two components (Table~\ref{tab:modelparms} and
Fig.~\ref{fig:w9models}). The uncertainties in the fitting parameters
are derived in the standard manner of fixing all but one parameter
which is varied until $\chi^2$ changes by unity.

\begin{table}
\caption[]{Best-fit parameters of the $r^{-2}$ shell model applied to
the continuum spectra of the compact and extended radio components of
W9.\label{tab:modelparms}}
\begin{tabular}{lllll}
\hline
Model &Component & $\dot M$~M$_\odot$yr$^{-1}$ & $R_i$ & $R_o$ \\
      &          & $(\times 10^{-5})$& (AU) & (AU)\\
\hline 
1 &Compact & $8.7^{+0.5}_{-0.2}$ & $0$ & $5800^{+7200}_{-1200}$ \\
2 &Extended& $17.6^{+0.5}_{-0.4}$ & $1100^{+80}_{-90}$ & $\infty$\\
\hline
3 &Compact & $9.2^{+0.4}_{-0.5}$ & $0$ & $4000^{+3000}_{-900}$ \\
3 &Extended & $33^{+12}_{-5}$ &$4000^{+3000}_{-900}$ & $\infty$\\
\hline
\end{tabular}

Notes: Models 1 and 2 correspond to best-fits of each individual
spectra. Model 3 assumes the outer radius of the compact component is
equal to the inner radius of the extended component.  For each model
it is assumed that $T_e=10$~kK, v$_\infty=200$~km\,s$^{-1}, Z=0.9,
\mu=1.5$, $\gamma=0.8$ and the volume filling factor
$f=1$.
\end{table}

A total of three models were examined.  Model~1 leads to a mass-loss
rate of $8.7^{+0.5}_{-0.2}\times10^{-5}({\rm v}_\infty/200~{\rm
km\,s}^{-1}) f^{1/2}$~M$_{\odot}$yr$^{-1}$ for $Z=0.9,
\mu=1.5,\gamma=0.8,T=10$~kK \citep[e.g.][]{Leitherer:1991} for a
volume filling factor for clumps $f$ \citep{Abbott:1981}. The
mass-loss rate is scaled to 200~km\,s$^{-1}$ since the outflow
velocity in W9 is not well constrained with observed line widths that
range between 40 and 800~km~s$^{-1}$ (CO8). Assuming the
same outflow velocity and clump structure, the derived mass-loss rate
of the extended region (Model 2) is approximately twice as high as the
current phase of mass-loss, implying an earlier era of higher
mass-loss rate $\sim10^{-4}$~M$_\odot$yr$^{-1}$.

The model spectrum of the individual radio components established that
the best-fit outer radius of the compact component (the current
stellar wind) is greater than the inner bound of the extended region
(earlier phase of mass-loss). If the mass-loss that produced both
components is radially symmetric, it is not plausible for the inner
radius of the extended region to be smaller than the extent of the
current stellar wind. The third model (Model 3) assumed the inner
bound of the extended region is at least as large as the outer bound
of the current stellar wind.  Fitting both continuum spectra
simultaneously through minimizing the total $\chi^2$ from the two
models gives the parameters of Model 3, which supports a factor of 3
higher mass-loss rate in the earlier phase of mass-loss than in the
current stellar wind.  The resulting model fits to the data are shown
in Fig.~\ref{fig:w9models}.

The derived mass-loss rate for the putative earlier mass-loss phase is
high for a massive star, being of the same order as the the current
mass-loss rate in $\eta$~Carina of $\approx 10^{-4} -
10^{-3}$~M$_{\odot}$ yr$^{-1}$ \citep{Corcoran:2008}. The inner bound
of $\sim4000$~AU suggests that this mass-loss epoch would have ended
$\sim200$~yrs ago assuming an outflow velocity of 200~km\,s$^{-1}$,
followed subsequently by the current mass-loss phase with $3\times$
lower density.

With such a model for the circumstellar environment, what does this
imply for the evolutionary status of W9? While veiling any
photospheric features, the rich optical-IR emission line spectrum and
IR excess indicate a rich circumstellar environment and permit a
formal sgB[e] star classification \citep[C08;][]{Ritchie:2010}. However
spectral peculiarities exist when compared to other examples; most
notably the unusual composite line profiles of He\,{\sc i} 5876 and
6678$\AA$.  The X-ray emission from W9 supports a binary system, being
too hard ($kT\sim3$~keV) and bright ($L_x\sim10^{33}$~erg\,s$^{-1}$)
to arise in a single star, although it does not constrain the nature
of the components \citep[C08;][]{Ritchie:2010}. Likewise, while W9 is
found to be photometrically variable, no period has been identified
currently \citep{Bonanos:2007}. Identification of W9 as a binary
awaits further supporting observational evidence.

The current mass-loss rate of W9 is approximately an order of
magnitude higher than that inferred for any of the other transitional
stars within Wd~1. It is directly comparable to those of the LBVs
\object{AFGL 2298}, \object{AG Car}, \object{FMM 362} and the Pistol
star in quiescence with $\dot M=3-6\times10^{-5}
{f^{1/2}}$~M$_{\odot}$yr$^{-1}$ \citep[][]{Clark:2009, Groh:2009,
Najarro:2009}, despite being a factor of $\sim2\times$ less
luminous\footnote{In the absence of a bolometric correction for W9
adopting a comparable luminosity to the cool hypergiants - appropriate
considering post-MS evolution is expected to proceed at constant
bolometric evolution prior to the WR phase - suggests
log(L/L$_{\odot}$)$\sim$5.8.}. Moreover, depending on the outflow
velocity and wind-clumping factor the mass-loss rate for W9 is
uncomfortably close to the limit expected for a line-driven wind for
which \cite{Smith:2006} estimate a maximum mass-loss rate of
\begin{displaymath}
1.4 \times10^{-4} (L/10^6L_{\odot})~{\rm M}_\odot\,{\rm yr}^{-1}. \nonumber
\end{displaymath} 
 
Such a high mass-loss rate and the presence of a detached shell of
material formed in a previous phase of enhanced mass loss is
suggestive of an LBV identification, with $\dot M \sim
3\times10^{-4}$~M$_{\odot}$yr$^{-1}$ for W9 during this event matching
the time averaged mass-loss rate inferred for the LBV Wra751 by
\cite{Voors:2000}, as well as the YHG IRC+10 420 during the formation
of its nebula (Sect. 4.3).  The observed line spectrum, absolute
visual magnitude and spectral energy distribution exclude the
possibility that W9 is currently in a cool hypergiant phase with the
binary companion providing the requisite ionizing flux to yield the
emission line spectrum. Therefore, if the enhanced mass-loss rate is
solely driven by the underlying star, this would be one of the few
known examples of an LBV caught in a high mass-loss rate
episode. Moreover, it would demonstrate that significantly enhanced
mass loss in $\sim40$M$_{\odot}$ stars is not solely limited to occurring in
a cool hypergiant phase.

However, while the current observational data are consistent with the
ejecta being formed as a result of the post-MS evolution of
a single massive star, its formation in an interacting binary system
undergoing significant mass transfer, in turn leading to common
envelope evolution, also appears viable and well motivated given the
evidence of W9 being a binary.  Examples of massive binaries with both
a normal and a relativistic companion are known: RY Scuti and SS433
respectively, and thus provide observational templates. As with W9,
both these systems have rich emission-line spectra and IR excesses due
to the presence of circumstellar dust \citep[][and references
therein]{Gehrz:1995, Clark:2007}. More interestingly, both are
associated with compact, bright radio nebulae with sizes and fluxes
directly comparable to W9 yet attributed to mass lost through binary
interaction \citep{Blundell:2001, Gehrz:1995}. The former authors
derive a mass-loss rate of the order of 10$^{-4}$~M$_{\odot}$yr$^{-1}$
from the radio observations for SS433, while the latter infer an
ionized mass of 10$^{-3}$~M$_{\odot}$ for the nebula associated with
RY Scuti, with a mass-loss rate of $\sim$10$^{-4}$ --
10$^{-5}$~M$_\odot$yr$^{-1}$ inferred from optical emission lines
\citep{deMartino:1992}; both rates are comparable with W9.

\subsection{The OB supergiants}
\label{sec:obsupergiants}
Currently, there are $\sim$75 spectroscopically identified evolved O
stars (O giants and OB supergiants) within Wd~1, with a comparable
number of stars classified as such on the basis of photometry
\citep[C05;~C08;][]{Negueruela:2010}\footnote{Note that we include
implicitly the mid-B supergiants such as W70, 71 and 57 here, while
deferring discussion of the more luminous, and likely more evolved, B
hypergiants W7, 13, 33 42a and 243 to
Section~\ref{sec:bhypergiants}.}.  Unfortunately, the currently
published spectroscopy of these objects is of insufficient resolution
and wavelength coverage to yield accurate physical
parameters. However, canonical wind parameters for late O/early B
supergiants (see C05, Table A2) indicate that at a distance of
$\sim4$~kpc thermal wind emission would be undetectable, due to a
combination of comparatively high terminal wind velocity and low
mass-loss rate. As an example, the O9.5 Ia star with v$_{\infty} \sim
1900$~km\,s$^{-1}$ and $\dot M=6\times10^{-6}$~M$_{\odot}$yr$^{-1}$
and adopting $Z=\gamma=1$ and $\mu=1.34$ \citep{Leitherer:1995} has an
expected 8.6-GHz flux of $\sim0.06$~mJy, much lower than the 8.6-GHz
detection limit of $0.17$~mJy. Since radio flux is $\propto f^{-3/2}$,
it would take a highly clumpy wind with $f\sim0.5$ to make such a wind
detectable.

Consequently, the apparent detections of W15 and W17 (O9Ib and O9Iab
respectively - Negueruela et al., in preparation), D09-R1 and
potentially D09-R2 (both B0I or earlier; C08) are
somewhat unexpected, most especially considering that these stars are
unremarkable compared with the rest of the OB stars studies
spectroscopically.  Of note is that W15, W17 and D09-R1 all are
coincident with extended emission regions and the parameters of the
unresolved radio components coincident with the underlying stars are
likely influenced by this extended emission. As a result, the derived
parameters may not reflect the intrinsic properties of the underlying
stars, but rather the extended plasma region.

That withstanding, W17 has a spectral index marginally steeper than
-0.1, hinting at non-thermal emission and a potential CWB. However, we
note it does not have detected X-ray emission C08 and
shows no evidence of RV variations \citep{Ritchie:2009a}. W15 has a
spectral index lower limit of 0.0, and it is not possible to
distinguish whether the emission is from either a stellar wind or a
non-thermal/thermal composite source.  It shows no evidence for
binarity in either RV observations \citep{Ritchie:2009a} or X-ray
observations (C08), where the emission is soft
($kT\sim0.6$~keV) and at a level expected for a single O-type star
\citep[$\sim2 \times10^{32}$ erg s$^{-1}$; e.g.][]{Berghoefer:1997} as
a result of shock heating within the stellar wind.  Nevertheless, if
the radio flux is attributed to thermal emission from a stellar wind,
W15 would have an unusually high mass-loss rate of $\dot
M=3\times10^{-5}$~M$_{\odot}$yr$^{-1}$ in comparison to other
late-O/early-B supergiants \citep{Crowther:2006,Mokiem:2007}. The
absence of RV variations in these stars does not rule out a binary
interpretation unequivocally since the mass ratios may be extreme and
the inclination of the orbit unfavorable for ready detection of RV
variations. Furthermore, the orbital period could be significantly in
excess of the observation timescale.

The spectral index of the point source associated with D09-R1 is
complicated by the presence of an extended component, and given the
absence of both optical spectroscopy and an X-ray detection, it is
simply concluded that the nature of this emission remains uncertain,
and note that if it is assumed to be a stellar wind, the same problems
with respect to an extreme mass-loss rate as found for W15 are
encountered. 

\subsection{The cool super- and hypergiants}
\label{sec:cool}

All four of the RSGs and at least four of the six YHGs in Wd~1 are
radio sources. Irrespective of the uncertain identification of W16a as
a radio emitter, the fraction of radio sources amongst both these
classes of object is high.  Of the RSGs, W75 was the weakest source
detected and was unresolved at any wavelength.  The other three
detected RSGs (W20, 26 and 237) have extended radio emission.

For both W20 and W26 the radio emission is best characterised by two
components: a compact ($\sim2\arcsec$), spatially resolved source
coincident with the star, and a more extended nebula
($>10\arcsec$). W237 is similar in structure though the inner
component coincident with the star is unresolved. In W20 and W26 the
more extended components have a distinct cometary morphology that
extends away from the central region of the cluster, while that in
W237 is more elliptical.

The radio properties of the YHGs are more heterogeneous. W8 is not
detected. W32 is associated with a weak point source, while W12 and
W265 are associated with compact resolved sources, similar to the RSG
radio sources. Finally, the radio emission from W4 consists of a point
source at the position of the star surrounded by an extended
component.

The majority of the cool super- and hypergiants have spectral
indices that are flat, within the uncertainties. Unlike OB and WR
stars, composite thermal and non-thermal spectra have not been
identified previously, and are not expected. Hence we identify these
indices with optically-thin thermal emission. For a Maxwellian plasma
at temperature $T$, the optically-thin radio flux density is given by
\begin{displaymath}
S_\nu=5.7\times10^{-56} T^{1/2}g_\nu D^{-2}E_V~~{\rm mJy},
\end{displaymath}
where $D$ is the distance in kpc and $E_V$ is the emission integral 
\begin{displaymath}
E_V=\int n_e^2\,dV=6\times10^{57}\bigl({D\over4~{\rm kpc}}\bigr)^2
\bigl({S_{8.6}\over1~{\rm mJy}}\bigr)\bigl({T\over10{\rm kK}}\bigr)^{-{1\over2}}~~{\rm cm}^{-3}
\end{displaymath}
at 8.6 GHz. The total ionized mass can then be determined from
$M_T=\mu m_H (E_V/n_e)$.

The ionized masses of the nebulae, based on the source parameters
determined at 8.6~GHz, are given in Table~\ref{tab:optthinparms}. Note
this is not the total mass since these observations are insensitive to
any neutral component. On average, the emission integral
$E_V\sim1.5\times10^{58}$~cm$^{-3}$, implies an ionizing
luminosity that has to be greater than $E_V\alpha^{(2)} \sim
6\times10^{45}$~s$^{-1}$, where $\alpha^{(2)}$ is the is the hydrogen
recombination rate to all but the ground state. Such ionizing
luminosity is not available from these relatively cool stars with
stellar effective temperatures $\leq5000$~K
\citep[e.g.][]{deJager:1998}.

How are these envelopes ionized? A cluster of mass $10^5$M$_\odot$
and solar metallicity can easily provide such a photon luminosity and
seems the most likely source of the ionizing radiation. Taking the
cluster to be approximately 1~pc across and with $\sim100$ OB-type
stars implies that on average each OB star occupies a volume
$0.04$~pc$^3$, and is separated from the next OB star by
$\sim0.34$~pc. If each OB star provides $10^{48}$~s$^{-1}$ ionizing
photons \citep{Martins:2005}, with $\sim10^{45}$~s$^{-1}$ required to
fully ionize the cool hypergiant envelopes, suggests a radiation
dilution factor $\propto (R_*/D)^{-2}$ less than 1000, which implies
the hypergiant envelopes are $\sim0.01$~pc radius. This is closely
consistent with the sizes observed for RSG nebulae e.g. VY CMa (see
Sec.~\ref{sec:redsuper}).

Alternatively, a hot companion star could provide the required photon
flux, which at $10^{45-46}$~s$^{-1}$ implies an early B-type
star. The YHG HR~8752 provides a precedent for such a model: it is
detected at radio wavelengths and together with [\ion{N}{ii}]-emission
lines, needs a source of ionizing radiation that is provided by a
putative B1 companion \citep{Stickland:1978}. In this respect we
highlight the presence of [\ion{N}{ii}]-emission lines in the optical
spectra of W20, W237 and W265 \citep{Clark:2010}.

If the plasma in these optically-thin nebulae is due to mass loss from
the underlying stars, having optically-thin spectra down to a
frequency of 1.4~GHz implies that the ionized extended envelopes have
inner radii significantly larger than the stellar radii. If this is
the case, two possible models for the envelope can be advanced. The
mass loss from these objects could have stopped or significantly
decreased at some point in the past resulting in a detached
shell. Alternatively, the ionized region may only occupy the periphery
of the envelope, with the interior part of the envelope being
neutral. Such a geometry could result from external ionization from
the cluster radiation field. High-resolution observations are required
to substantiate either of these potential models.

\begin{table}
\caption[]{Ionized mass estimates for the optically-thin stellar nebulae
 around the YHGs (top panel) and RSGs (bottom panel) in Wd~1.}
\label{tab:optthinparms}
\begin{tabular}{llllllll}
\hline
 ID  & S$_{8.6}$ & Maj & ${{\rm Min}\over{\rm Major}}$ &  Vol &   $E_V$ &  $n_e$ &   Mass \\
     &  & & &          $10^{50}$& $10^{58}$&  $10^3$  &  $10^{-3}$\\
     & (mJy)  & ('') && (cm$^3$) &  (cm$^{-3}$) & (cm$^{-3}$) & (M$_\odot$) \\
\hline
 4   &  2.2 & 2.45 &0.49 &   4.0&   1.3&  5.7&   2.9\\
 12a &  2.9 & 1.82 &0.92 &   5.7&   1.7&  5.5&   3.9\\
  265 &  2.3 & 2.86 &0.54 &   7.7&   1.4&  4.2&   4.1\\
\hline
  20  & 15.6  &13.79 & 0.48 & 677.9 & 9.2 & 1.2 & 99.6 \\
  26  & 173.4 &15.0  &0.33 & 412.4& 102.2 &  5.0& 259.0\\
 237    &  5.6 &11.19 &0.76 & 908.0&   3.3&  0.6&  69.1\\
\hline
\end{tabular}
Note: A distance of 4~kpc is assumed

\end{table}

%

\subsubsection{The Yellow Hypergiants}
\label{sec:yhg}
Given the rarity of YHGs and the difficulty identifying suitable
observational diagnostics, few measurements of the quiescent or
outburst wind velocities and mass-loss rates of YHGs exist in the
literature. For \object{HD 33579} (A3Ia$^+$) in the LMC,
\cite{Stahl:1991} quote v$_\infty \sim200$~km\,s$^{-1}$ and $\dot
M\sim2\times10^{-6}$~M$_{\odot}$yr$^{-1}$, and \cite{Israelian:1999}
report v$_\infty\sim$120km\,s$^{-1}$ and $\dot
M<6.7\times10^{-6}$M$_\odot$yr$^{-1}$ for \object{HR 8752} in
1998. For the well-known example \object{$\rho$~Cas}, \cite{Lobel:1998} find
v$_\infty\sim120$~km\,s$^{-1}$ and $\dot
M<9.2\times10^{-5}$M$_{\odot}$~yr$^{-1}$ in 1993, in contrast to
v$_\infty\sim35$~km~s$^{-1}$ and $\dot
M\sim5.4\times10^{-2}$~M$_\odot$yr$^{-1}$ during the 2000 outburst
\citep{Lobel:2003}.  \cite{Castro-Carrizo:2007} measure outflow
velocities of 35~km\,s$^{-1}$ and $25-37$~km\,s$^{-1}$ for the ejecta
associated with \object{HD179821} and \object{IRC +10 420} (although
{\em a priori} it is not clear if they were ejected in a RSG or YHG
phase).  For both these stars, variable mass-loss rates of
$4-300\times10^{-5}$~M$_\odot$yr$^{-1}$ and
$1.2-3\times10^{-4}$~M$_{\odot}$yr$^{-1}$ were found respectively
during the nebulae formation events, with \cite{Blocker:1999} also
reporting a mass-loss rate in excess of $10^{-3}$~M$_{\odot}$yr$^{-1}$
for IRC +10 420 in the recent past. These works all suggest that
during outburst the mass-loss rates in YHGs are higher than during
quiescence whereas the outflow velocity is lower during outburst.

Taking a mean 8.6-GHz flux of 2.5~mJy for the YHGs, their radio
luminosity is $4.9\times10^{19}$~erg~s$^{-1}$. The only other radio
detected YHG in the Galaxy, HR\,8752, has an 8.6-GHz flux of 17~mJy
(Dougherty et al., in preparation) at a distance of 1.7~kpc, giving a
luminosity of $5.9\times10^{19}$~erg~s$^{-1}$, similar to the YHGs in
Wd~1. HR~8752 has been interpreted to have a stellar wind from
measured spectral indices of $0.51\pm0.08$ \citep{Higgs:1978} and
$0.61\pm0.04$ (Dougherty et al., in preparation). Only the compact
component in W4 appears to have a spectral index consistent with a
stellar wind ($+0.49\pm0.31$), but given the uncertainty, arguably
also consistent with optically-thin thermal emission. Assuming $Z=0.9,
\mu=1.5$ and $\gamma=0.8$ as for W9 and that the putative wind from W4
is fully ionized and smooth $(f=1)$ gives a mass-loss rate of $0.7
\times 10^{-5}({\rm v}_{\infty}/200{\rm
km\,s}^{-1})$~M$_{\odot}$yr$^{-1}$. This is consistent within a factor
of a few with the quiescent mass-loss rates of \object{HD 33579},
\object{HR 8752} and \object{$\rho$~Cas}, but significantly lower than
inferred for \object{$\rho$~Cas}, \object{HD179821} and \object{IRC
+10 420} during outburst, especially for comparably low velocities
($<40$km\,s$^{-1}$).

The extended nebulae around W4, W12a and W265 have masses of
$3-4\times10^{-3}$~M$_{\odot}$ and radii of $\sim0.02$~pc (based on
the major axis size in Table~\ref{tab:optthinparms}). An assumed
outflow velocity of $\sim200$~km\,s$^{-1}$ gives a flow time of
$\sim100$~yrs, during which time mass loss at a rate of
$\sim10^{-5}$~M$_\odot$yr$^{-1}$, as found for W4 above, yields a
total mass loss of $10^{-3}$~M$_{\odot}$. Thus, by comparison with the
cases of \object{$\rho$~Cas}, \object{IRC +10 420} and \object{HR
8752}, the radio nebulae associated with the YHGs in Wd~1 may be the
result of quiescent mass loss, rather than an outburst phase.

None of the YHGs in Wd~1 possess the old ($\sim4000$~yr), massive
(M$>1$~M$_{\odot}$) ejection nebulae seen in HD179821 and IRC +10 420
\citep{Oudmaijer:1996, Castro-Carrizo:2007}. We suspect that in the
extreme environment of Wd~1 such nebulae would quickly be ablated by
the cluster wind arising from the OB stars - as appears to be the case
for the RSGs W20 and W26 (see Sec.~\ref{sec:redsuper}). However we
cannot exclude the possibility that the YHGs may be evolving to cooler
temperatures (redwards) and have yet to encounter the instabilities
leading to extreme mass-loss rates in either the RSG or post-RSG
phase. A mixture of both pre- and post-RSG objects combined with the
intrinsic variability of YHGs \citep{Clark:2010} may explain the
heterogeneous radio properties of the Wd~1 population. This argument
is supported by evidence that even the best two candidates for
post-RSG stars, HD179821 and IRC +10 420, exhibit quite dissimilar
nebular properties \citep{Oudmaijer:2009, Patel:2008}.

\subsubsection{The Red Supergiants}
\label{sec:redsuper}

Given very high mass-loss rates inferred for extremely luminous RSGs
(log(L/L$_{\odot}$)$>5.5$), it is suspected that they play an
important role in post-MS evolution of stars at the Humphreys-Davidson
limit. Consequently, they have been the subject of numerous
(multi-wavelength) studies to constrain their mass-loss histories. Of
the Galactic examples, \cite{Schuster:2006} used the HST/WFPC2 to
identify compact (${\rm r}\leq0.003$pc) nebulae associated with \object{NML
Cyg}, \object{VX Sgr} and \object{S Per}, with \cite{Morris:1983} also
associating a much larger ($>$1pc) \ion{H}{ii} region with
\object{NML Cyg}.  \cite{Dewit:2008} report the presence a nebula
around \object{$\mu$ Cep} at 25~microns (${\rm r}\sim0.03$~pc); a comparably
sized nebulae surrounds \object{VY CMa} \citep[${\rm r}\sim0.04$~pc in
HST/WFPC2 images;][]{Smith:2001}. Finally \cite{Serabyn:1991} and
{\cite{Yusef-Zadeh:1991} report the presence of an asymmetric nebula
($\sim$1pc in length) around the Galactic Centre RSG \object{IRS7} at both
near-IR and radio wavelengths.

As with these stars, the discovery of large nebulae associated with
three of the 4 RSGs in Wd~1 (major axis $>0.2$~pc;
Table~\ref{tab:optthinparms}) emphasises the role of mass loss in this
evolutionary stage. \object{NML Cyg} and \object{IRS7} are of special
interest in the context of the RSGs in Wd~1, given that these stars
are also located within dense stellar aggregates, namely Cyg OB2
\citep{Morris:1983} and the Galactic Centre cluster
\citep[e.g.][]{Yusef-Zadeh:1991} respectively. Indeed the pronounced
cometary nebulae of W20 and W26 appear similar both in morphology and
linear scale to that of IRS7.  Consequently we attribute the shape
of these nebulae to a similar physical process, namely the interaction
of the ejecta/RSG stellar wind with the cluster medium/wind. Lacking
dynamical information, we refrain from attempting a quantitative
analysis of the nebular morphologies, but note that the unremarkable
radial velocities found for these stars by \cite{Mengel:2007} suggest
that the dominant cause of ram pressure shaping the nebulae is the
expansion of the cluster wind, rather than the rapid motion of the
stars through a quasi-static cluster medium. This is supported by the
orientation of each of these cometary nebulae away from the cluster
centre region.

The morphology of the radio nebula associated with W237 suggests it
has been less affected by the cluster wind than either W20 or W26.
Adopting an outflow velocity of 30~km\,s$^{-1}$ for the RSG wind, the
mass of 0.07~M$_{\odot}$ and radius of $\sim0.11$~pc
(Table~\ref{tab:optthinparms}) results in a kinematic age of 3,600~yr
and a time averaged mass-loss rate of
2$\times$10$^{-5}$~M$_{\odot}$yr$^{-1}$. This age supports the
suggestion that the radio nebula is little affected by the cluster
wind, otherwise it likely would be ablated as observed in W20 and
W26. This is consistent with its position further from the cluster
centre than either W20 or W26. The derived mass-loss rate is
comparable to the mass-loss rates found for field RSGs, which range
from 10$^{-6}$ - 10$^{-4}$~M$_{\odot}$yr$^{-1}$ \citep{Jura:1990,
Sylvester:1998}, although transient rates as high as
$\sim$10$^{-3}$~M$_{\odot}$yr$^{-1}$ (\object{NML Cyg} and \object{VY
Cma}; \citep{Blocker:2001,Smith:2001} and as low as a $\sim$few
10$^{-7}$~M$_{\odot}$ in \object{$\mu$ Cep} \citep{Dewit:2008} have
been observed.

Moreover, both W20 and W26 have similar nebular masses to W237 and, in
turn, all three are directly comparable to that of the nebula
surrounding VY CMa \citep{Smith:2001}. Consequently the ejecta
associated with the RSGs within Wd~1 appear to be the result of
similarly extreme, possibly transient mass-loss events as those that
yield the nebulae around other Galactic RSGs
\citep[e.g.][]{Smith:2001}. In this respect, the apparent
instabilities in W26 and W237 are of interest \citep{Clark:2010}.

\subsection{The B hypergiants}
\label{sec:bhypergiants}
The final subset of transitional stars to be considered are the B
hypergiants, which likely represent the immediate progenitors of
WN-type stars. Included in this group are the LBV W243 and the cool
B-type hypergiants W7, 33 and 42. The final member is the early-B
hypergiant W13, which emphasises the close physical continuity between
the B hypergiants and the hydrogen-rich late-WN stars such as W44
(=WR L (WN9h); C08) detected at 8.6GHz
(Table~\ref{tab:radiometry} and Sec~\ref{sec:wr}). Of these stars, only
W243 has a spectral index indicative of thermal wind emission.

W243 has recently been the subject of an extensive spectroscopic study
by \cite{Ritchie:2009b}. This confirmed the finding of
C02 that W243 had substantially cooled since the
previous observations of \cite[e.g.][]{Westerlund:1987}, with the
stellar temperature between 2002-8 being 8-9kK, with a formal spectral
classification of mid-A type and a photospheric absorption spectrum of
a YHG. However, detailed non-LTE analysis were unable to
simultaneously reproduce these features and the prominent H\,{\sc i}
and He\,{\sc i} emission lines, leading \cite{Ritchie:2009b} to
conclude that W243 harbored a hot binary companion. Unfortunately,
current spectroscopic data are unable to constrain the nature of the
hot companion; nevertheless at this time W243 closely resembles the
known YHG binary HR8752 (Sec.~\ref{sec:yhg}) and we assume that the
hot companion is responsible for the ionization of its wind.

Given the {\em current} properties of W243 we adopt identical values
of Z, $\mu$ and $\gamma$ as used for the YHG W4, leading to a
mass-loss rate of $1.1 \times 10^{-5} ({\rm v}_\infty/200{\rm
km\,s}^{-1}){f}^{1/2}~{\rm M}_\odot\,{\rm yr}^{-1}$, which is
directly comparable to W4. This is to be expected considering their
similar radio fluxes (Table~\ref{tab:radiometry}). This mass-loss rate
is an order of magnitude higher than that found by
\cite{Ritchie:2009b}, correcting for difference in assumed outflow
velocity, though it should be noted that the observed optical spectrum
is relatively insensitive to the current mass-loss rate. Assuming a
4$\sigma$-detection limit of 0.24~mJy for the remaining stars yields a
corresponding upper limit to the mass-loss rates of $2\times
10^{-6}({\rm v}_\infty/200{\rm km\,s}^{-1})~{\rm M}_\odot\,{\rm
yr}^{-1}$.

In comparison, the LBVs \object{HR Car} and \object{HD 160529} have
mass-loss rates of $\sim1-2\times10^{-5}f^{1/2}$~M$_{\odot}$yr$^{-1}$
determined from radio observations, with \object{HD 80077} and
\object{HD 168607} an $\sim$order of magnitude lower
\citep{Leitherer:1995}.  Therefore, we conclude that the mass-loss
behavior of the Wd~1 population of B hypergiants is consistent with
expectations drawn from the wider Galactic population.  No extended
emission is associated with any of these stars, whereas a number of
Galactic LBVs are associated with radio bright nebulae. However such
nebulae are typically found to be old \citep[$\sim
10^3-10^4$~yr][]{Clark:2003} and of an extent sufficient to encompass
the central regions of Wd~1. As argued for the YHGs, we suspect that
they would quickly be disrupted by the cluster wind and/or radiation
field.

\subsection{Wolf-Rayet stars}
\label{sec:wr}
In recent years, Wd~1 has been shown to harbour at least 24 WR stars
\citep{Crowther:2006}, representing 8\% of the currently known
galactic population. Five of these have been detected in this radio
survey at a flux level typically $0.3-0.5$~mJy at 8.6~GHz: WR~A
(=W72), B, F(=W239), L(=W44), and V(=W31b) \citep[][and
references therein]{Crowther:2006}, {\em potentially} consistent with
the radio emission originating in their stellar winds. 

WR~B is an exception, with a deduced flux of 4.3 mJy. This flux
is $5-10\times$ higher than the other detected WR stars, implying a
wind density that is $3-6\times$ higher than the mean value for the
other WRs.  \cite{Bonanos:2007} found that WR~B is an eclipsing binary
and in binary systems the flux can be raised substantially above that
from a stellar wind by emission arising in a wind-collision region
(WCR) from the colliding winds of two massive stars in a binary
\citep[e.g][]{Dougherty:2005}. More likely, the flux of WR~B is biased
by being a part of a more extended emission region (A3;
Sec.~\ref{sec:extended}), a suggestion supported by data reduction
that indicates the 8.6-GHz emission is resolved which is not expected
for typical WR winds at this distance.  For comparison, the closest WR
star to the Sun, WR\,11 at $\sim$0.26~kpc \citep{vanderHucht:1997} has
a stellar wind that is resolved by the ATCA at 8.6 GHz, with a size of
0.47 arcseconds (Dougherty et al., in prep). If WR\,11 was at a
distance of 4~kpc, the wind would be $\sim30$~mas in diameter and not
resolved. This leads us to believe that the source parameters derived
for WR~B are influenced by its location in the extended emission
region A3.

W44 (WR~L) is the only WR star that has a spectral index consistent
with a partially optically-thick stellar wind, although the lower
limit to the spectral indices of WR F leave this possibility open. For
the WNVLh star WR~L we adopt $\mu$=2.0, Z=1.0 and $\gamma$=1.0
\citep[e.g.][]{Leitherer:1997}, resulting in $\dot M =2\times
10^{-5}({\rm v}_{\infty}/1000{\rm km\,s}^{-1})
f^{1/2}$~M$_{\odot}$yr$^{-1}$ respectively. For the WC9 star WR~F
adopting $\mu$=4.7, Z=1.1 and $\gamma$=1.1 yields $\dot M=3.3\times
10^{-5} ({\rm v}_{\infty}/1000{\rm
km\,s}^{-1})f^{1/2}$~M$_{\odot}$yr$^{-1}$. Both these mass-loss rates
are comparable to those found for similar stars in the Galactic Centre
cluster by \cite{Martins:2007} and other field WRs
\citep[e.g.][]{Cappa:2004}.

The remaining three WR detections (WR~A, F, V) have flat, though
highly uncertain, spectral indices. If they are flat, an
interpretation due to composite spectra composed of both thermal and
non-thermal emission components would be consistent with that observed
in a number of WR+OB binaries \citep{Dougherty:2000}. Certainly, among
WR stars with flat radio spectra, those that have been observed at
high resolution \citep[e.g.][and references therein]{Williams:1997,
Dougherty:2000, Dougherty:2005}, all have identified WCRs arising in
CWBs.

Recently, Wd~1 has been observed by CHANDRA \citep{Clark:2007,
Skinner:2006, Muno:2006a, Muno:2006b} and the WR stars WR\,A, B, F and
L are all X-ray bright ($L_x\sim10^{32-33}$ erg~s$^{-1}$) and hard
($kT>2.6$ keV i.e.  $T>3\times10^7$~K). These temperatures are
expected for plasma heating by shocks in a WCR \citep{Stevens:1992,
Pittard:2009a, Pittard:2009b}, which led both \cite{Clark:2007} and
\cite{Skinner:2006} to suggest these WR systems are all 
CWB systems. In support of this argument, each of these WR
stars have exhibited photometric variations in the optical
\citep{Bonanos:2007}. WR\,B demonstrates clear eclipses while WR\,A
demonstrates a 7.63-day periodic modulation to its light curve and is
also found to be a spectroscopic binary \citep{Crowther:2006}, though
such a short-period system is not anticipated to exhibit evidence of
non-thermal radio emission due to a wind collision
\citep{Dougherty:2000}. However, if all these X-ray bright WR stars
are interpreted as CWBs, then the apparently thermal radio spectrum of
WR~L would be attributed to a sufficiently short-period binary where
any non-thermal radio emission from a WCR is completely absorbed by
the circumbinary stellar wind(s) or the relativistic electrons are
cooled by the intense UV radiation of a putative massive companion.

The IR excess associated with WR F has been attributed to hot dust
\citep{Crowther:2006, Groh:2006}, again supporting a CWB
interpretation, since dusty WC stars are widely considered to be in
binaries with OB companions where wind-wind interaction is the essential
mechanism that attains the high gas densities required for dust
formation \citep[e.g.][]{Williams:1999, Tuthill:1999,
Monnier:1999, Williams:2009}

A synthesis of the IR, X-ray and these radio observations therefore
argues for a high binary fraction amongst the WR stars, with
C08 suggesting that at least 70\%, and possibly the
complete population, of WR stars in Wd~1 are in massive binary
systems. However, given the lack of radial velocity variations from
spectral line observations this assertion has yet to be verified, and
may be exceedingly challenging to verify.

\section{Extended emission}
\label{sec:diffuse}}
Lastly, we consider briefly the extended emission and its origin.  Trivially,
optically-thin thermal emission is associated with young massive
clusters still embedded in Giant H\,{\sc ii} regions, where the young
massive stars ionize the remnants of the natal Giant Molecular
Cloud. However, this seems unlikely for Wd~1. At an age of $\sim$5Myr
one might expect the action of a cluster wind to have sculpted a
cavity $\sim10$~parsec in extent e.g. the G305 star forming complex
surrounding the young ($\geq$2-5Myr) massive clusters Danks 1 and Danks 2;
\cite{Clark:2004}, while the optically-thin thermal clumps associated
with Wd~1 are found significantly closer to, and indeed within the
confines of, the cluster. \cite{Kothes:2007} report the presence of a
large (25pc radius) H\,{\sc i} bubble (their feature B3) in the
neutral environment of Wd~1, which they attribute to the action of the
cluster wind on the natal GMC and wider ISM. If the cluster wind swept
the region clear of the natal material, the diffuse emission would
then have to arise from material ejected recently from the stellar
population of Wd~1, either via stellar winds or SNe. Assuming
velocities of 20-200km\,s$^{-1}$ (appropriate for RSG/LBV and SNe
ejecta) at a displacement of a $\sim$parsec yields an age for material
of 5$\times$10$^{3-4}$yr. Furthermore, a total ionized mass of
15~M$_\odot$ only requires an average mass-loss rate of a
$3\times10^{-6}$~M$_\odot$yr$^{-1}$ over the cluster lifetime, readily
available from the population in Wd~1 as discussed in the previous
sections.

An origin in recent stellar ejecta seems to be certain for stars such
as W20 and W26 with their large cometary nebula, but may not be so
clear for other regions such as A6, where no associated stellar source
is readily apparent in the FORS2 image. The ionized mass of A6 is
estimated to be $\sim0.5$~M$_\odot$, which could have certainly
originated from a stellar wind. In addition, A6 does have a
cometary-like structure that also appears to point towards the cluster
centre region, like the nebulae around W20 and W26 (see
Fig.~\ref{fig:dustoverlay}). An alternative explanation for some of
the diffuse emission is that it arises in natal material left behind
when the cluster region was largely swept clear by the cluster wind,
possibly through shielding by over-dense regions in the natal
cloud. The nature and origins of the extended emission will be
explored further in a forthcoming paper.

\section{Summary}
\label{sec:summary}

We have presented the results of a multi-frequency radio survey of the
galactic Super Star Cluster Wd~1, and used optical, IR and X-ray
observations from the literature to elucidate the nature of the radio
sources detected. We detect 18 radio sources for which positional
coincidence suggests an association with a cluster member.  These
radio stars are associated with every class of post-MS star present in
the cluster. Moreover, they comprise a diverse population of
point-like, unresolved sources and extended, resolved sources, with
spectral indices corresponding to thermal, non-thermal and composite
thermal and non-thermal emission. Nevertheless, it appears possible to
understand these properties under the current evolutionary paradigm
for massive (binary) stars. In brief, the radio observations reveal
the following properties for the massive stars in Wd~1:

\begin{itemize}
\item{} The brightest stellar radio source in Wd~1 is W9, with an
8.6-GHz luminosity that places it amongst the most luminous radio
stars known. The emission comprises both a point source and a more
extended nebula. The flux and spectral index of the point source
implies an origin in a powerful stellar wind, with a mass-loss rate of
$9.2^{+0.4}_{-0.5}\times 10^{-5} ({\rm v}_{\infty}/200{\rm
km\,s}^{-1}) f^{1/2}$~M$_{\odot}$ yr$^{-1}$.  The extended nebula is
deduced to arise from an earlier mass-loss epoch with a mass-loss rate
$\sim$3 times higher than the current stellar wind, close to the limit
expected for a line-driven wind \citep{Smith:2006}. An obvious
comparison to make is to LBVs, with W9 having a current mass-loss rate
similar to galactic examples e.g. the Pistol star. Likewise, the
mass-loss rate deduced for the extended nebula is comparable to
several other galactic LBV's during outburst, although orders of
magnitude less than inferred for both P Cygni and $\eta$ Carinae
during outburst \citep[][and references therein]{Clark:2009}.  The
X-ray properties of W9 imply that it is a binary system, although it
is not currently possible to constrain the nature of the companion. An
alternative scenario with significant mass loss in a common envelope
phase of stellar evolution cannot be excluded.

\item{} Surprisingly, three of the $>$100 evolved OB stars (luminosity
classes III-Ia) are detected, with radio fluxes an order of magnitude
larger than expected for stellar winds in these types. It is suggested
they may be CWBs based on potentially composite spectra of both
non-thermal and thermal emission. However, none of the three detected
stars exhibit X-ray emission characteristic of CWBs or evidence of RV
variations due to a binary companion. Alternatively, their seemingly
high radio luminosity may be due the influence of the extended
emission in which they are embedded.

\item{} All four RSGs are detected, with three associated with large
nebulae with ionized masses up to $\sim0.26$~M$_{\odot}$, emphasising
the importance of mass loss in this evolutionary phase. Of these, the
nebulae around W20 and W26 have a pronounced cometary morphology,
suggesting significant interaction with either the intracluster medium
or cluster wind. W237 shows less evidence for such interaction and has
a kinematic age of $\sim$3,600~yr and a {\em time averaged} mass-loss
rate of $2\times10^{-5}({\rm v}_{\infty}/30{\rm
km\,s}^{-1})$~M$_{\odot}$yr$^{-1}$. This is consistent with other
field RSGs, although it is substantially lower than inferred for
NML~Cyg and VY~CMa during the formation of their nebulae.

\item{} The YHG W4 is argued to have a stellar wind with a mass-loss
rate of $10^{-5}($v$_{\infty}/200{\rm
km\,s}^{-1})f^{1/2}$~M$_{\odot}$yr$^{-1}$, consistent with the few
estimates available for other field YHGs. The extended nebulae
associated with W4, 12a and 265 are significantly less massive than
those associated with the RSGs in Wd~1. It is argued they arise from
quiescent mass loss rather than during outburst episodes.

\item{} Since neither the YHGs nor RSGs are hot enough to ionize their
own stellar winds and/or more extended nebulae, the requisite ionizing
photons must arise from either an unseen companion or the cluster
radiation field.

\item{} Of the extreme B-type hypergiants, only the LBV W243 was
detected, with a spectral index consistent with thermal emission.  The
corresponding mass-loss rate is directly comparable to that found for
the YHG W4, as expected given the similarity in current spectral type
and radio flux. Upper limits of $2\times10^{-6}($v$_{\infty}/200{\rm
km\,s}^{-1})f^{1/2}$~M$_{\odot}$yr$^{-1}$ for the three other B
hypergiants were found, consistent with mass-loss rates amongst field
stars of these types\citep[][]{Leitherer:1995}.

\item{} Five of the 24 WRs known in Wd~1 were detected. Of these, WR L
has a partially optically-thick wind, with a mass-loss rate consistent
with stars of identical spectral type in the Galactic Centre cluster
and the general field population.  The remaining three (WR~A, B and V)
are identified as having composite spectra from a CWB. The optical and
X-ray properties of WR A and WR B have previously indicated these to
be binaries, while this is the first hint of binarity for WR V.
\end{itemize}

In combination with X-ray, optical and near-IR datasets, these radio
observations of Wd~1 provide striking evidence for the prevalence of
binarity among the massive stars in Wd~1. This lends strong support to
the previous estimates of high binary fractions, in excess of $>$40\%
\citep[C08,][]{Ritchie:2009a}.

The rich stellar population of Wd~1 permits us to investigate the
evolution of mass-loss rates as stars evolve from the MS.  With the
reduction of O-star mass-loss rates due to wind clumping
\citep[e.g.][]{Repolust:2004}, it has been recognised that the
majority of mass loss must occur in the post-MS transitional
phase. While the exact evolutionary sequence through the transitional
`zoo' is uncertain for stars of M$_{initial}$$\sim$40~M$_{\odot}$, it
is interesting that the radio mass-loss rates directly determined for
the YHG W4, the LBV W243 and the WNLh star WR L, which form a direct
evolutionary sequence in some schemes, are all closely equal to one
another.

However, with the exception of W9, such mass-loss rates
($\leq$10$^{-5}$M$_{\odot}$yr$^{-1}$) are likely insufficient to
remove the H-rich mantle unless stars remain in the transitional phase
for significantly longer than expected. This in turn suggests that an
additional mechanism is required to shed the requisite mass, with
short-lived episodes of greatly enhanced mass loss an obvious
candidate. Indeed mass-loss rates $>10^{-4}$M$_{\odot}$yr$^{-1}$ have
already been inferred for RSGs \citep[e.g. VY CMa;][]{Smith:2001} and
directly observed for the YHG $\rho$ Cas \citep{Lobel:2003}. The
nebulae around the RSGs W20, 26 and 236 already indicate that
significant mass loss has occurred for some stars within Wd~1, while
the mass-loss rate inferred for W9, over a magnitude greater than any
other transitional star, is of obvious interest. Indeed, is W9
undergoing an `eruptive' event currently?

We note that both the optical spectrum and spectral energy
distribution of W9 indicate that it is not a cool hypergiant,
suggesting that extreme mass-loss events are likely not confined to a
single evolutionary state, such as RSGs, and instead occur in both hot
and cool transitional phases. Indeed long term spectroscopic
observations indicate significant instability in the early-B
hypergiant/WNVL, RSG and YHG populations \citep{Clark:2010}.  Hence
one may not {\em a priori} assume a single physical mechanism leads to
enhanced mass loss in post-MS stars. When combined with detailed
modeling of the stellar spectra, further observations to derive the
nebular properties of cluster members, such as expansion velocity and
chemical abundances, would be very valuable in order to constrain the
phases in which enhanced mass loss occurs; thus Wd~1 will be
invaluable for investigating the physics that mediates the passage of
MS O-type stars to the Wolf-Rayet phase.

\begin{acknowledgements}
We thank Paul Crowther, Ben Davies, Simon Goodwin, Rene
Oudmaijer, Julian Pittard, Ben Ritchie and Rens Waters for many
stimulating discussions related to this work and for providing
numerous comments on early versions of the manuscript. A special
thanks to Rob Reid for advice on using the {\sc smerf} patch to {\sc
difmap} and to Bob Sault for his extensive advise on the use of {\sc
miriad}. The Australia Telescope Compact Array is part of the
Australia Telescope which is funded by the Commonwealth of Australia
for operation as a National Facility managed by CSIRO. SMD would like
to thank the Anton Pannekoek Institute and Open University or their
hospitality during a number of visits.  JSC is supported by a UK
Research Council (RCUK) Fellowship. This research is partially
supported by the Spanish Ministerio de Ciencia e Innovaci\'on (MICINN)
under grants AYA2008-06166-C03-03 and Consolider-GTC CSD2006-70.
\end{acknowledgements}


\begin{thebibliography}{96}
\expandafter\ifx\csname natexlab\endcsname\relax\def\natexlab#1{#1}\fi

\bibitem[{{Abbott} {et~al.}(1981){Abbott}, {Bieging}, \&
  {Churchwell}}]{Abbott:1981}
{Abbott}, D.~C., {Bieging}, J.~H., \& {Churchwell}, E. 1981, \apj, 250, 645

\bibitem[{{Benjamin} {et~al.}(2003){Benjamin}, {Churchwell}, {Babler}, {Bania},
  {Clemens}, {Cohen}, {Dickey}, {Indebetouw}, {Jackson}, {Kobulnicky},
  {Lazarian}, {Marston}, {Mathis}, {Meade}, {Seager}, {Stolovy}, {Watson},
  {Whitney}, {Wolff}, \& {Wolfire}}]{Benjamin:2003}
{Benjamin}, R.~A., {Churchwell}, E., {Babler}, B.~L., {et~al.} 2003, \pasp,
  115, 953

\bibitem[{{Berghoefer} {et~al.}(1997){Berghoefer}, {Schmitt}, {Danner}, \&
  {Cassinelli}}]{Berghoefer:1997}
{Berghoefer}, T.~W., {Schmitt}, J.~H.~M.~M., {Danner}, R., \& {Cassinelli},
  J.~P. 1997, \aap, 322, 167

\bibitem[{{Bl{\"o}cker} {et~al.}(1999){Bl{\"o}cker}, {Balega}, {Hofmann},
  {Lichtenth{\"a}ler}, {Osterbart}, \& {Weigelt}}]{Blocker:1999}
{Bl{\"o}cker}, T., {Balega}, Y., {Hofmann}, K.-H., {et~al.} 1999, \aap, 348,
  805

\bibitem[{{Bl{\"o}cker} {et~al.}(2001){Bl{\"o}cker}, {Balega}, {Hofmann}, \&
  {Weigelt}}]{Blocker:2001}
{Bl{\"o}cker}, T., {Balega}, Y., {Hofmann}, K.-H., \& {Weigelt}, G. 2001, \aap,
  369, 142

\bibitem[{{Blundell} {et~al.}(2001){Blundell}, {Mioduszewski}, {Muxlow},
  {Podsiadlowski}, \& {Rupen}}]{Blundell:2001}
{Blundell}, K.~M., {Mioduszewski}, A.~J., {Muxlow}, T.~W.~B., {Podsiadlowski},
  P., \& {Rupen}, M.~P. 2001, \apjl, 562, L79

\bibitem[{{Bonanos}(2007)}]{Bonanos:2007}
{Bonanos}, A.~Z. 2007, \aj, 133, 2696

\bibitem[{{Borgman} {et~al.}(1970){Borgman}, {Koornneef}, \&
  {Slingerland}}]{Borgman:1970}
{Borgman}, J., {Koornneef}, J., \& {Slingerland}, J. 1970, \aap, 4, 248

\bibitem[{{Brandner} {et~al.}(2008){Brandner}, {Clark}, {Stolte}, {Waters},
  {Negueruela}, \& {Goodwin}}]{Brandner:2008}
{Brandner}, W., {Clark}, J.~S., {Stolte}, A., {et~al.} 2008, \aap, 478, 137

\bibitem[{{Cappa} {et~al.}(2004){Cappa}, {Goss}, \& {van der
  Hucht}}]{Cappa:2004}
{Cappa}, C., {Goss}, W.~M., \& {van der Hucht}, K.~A. 2004, \aj, 127, 2885

\bibitem[{{Castro-Carrizo} {et~al.}(2007){Castro-Carrizo}, {Quintana-Lacaci},
  {Bujarrabal}, {Neri}, \& {Alcolea}}]{Castro-Carrizo:2007}
{Castro-Carrizo}, A., {Quintana-Lacaci}, G., {Bujarrabal}, V., {Neri}, R., \&
  {Alcolea}, J. 2007, \aap, 465, 457

\bibitem[{{Chapman} {et~al.}(1999){Chapman}, {Leitherer}, {Koribalski},
  {Bouter}, \& {Storey}}]{Chapman:1999}
{Chapman}, J.~M., {Leitherer}, C., {Koribalski}, B., {Bouter}, R., \& {Storey},
  M. 1999, \apj, 518, 890

\bibitem[{{Clark} {et~al.}(2007){Clark}, {Barnes}, \& {Charles}}]{Clark:2007}
{Clark}, J.~S., {Barnes}, A.~D., \& {Charles}, P.~A. 2007, \mnras, 380, 263

\bibitem[{{Clark} {et~al.}(2009){Clark}, {Crowther}, {Larionov}, {Steele},
  {Ritchie}, \& {Arkharov}}]{Clark:2009}
{Clark}, J.~S., {Crowther}, P.~A., {Larionov}, V.~M., {et~al.} 2009, \aap, 507,
  1555

\bibitem[{{Clark} {et~al.}(2003){Clark}, {Egan}, {Crowther}, {Mizuno},
  {Larionov}, \& {Arkharov}}]{Clark:2003}
{Clark}, J.~S., {Egan}, M.~P., {Crowther}, P.~A., {et~al.} 2003, \aap, 412, 185

\bibitem[{{Clark} {et~al.}(1998){Clark}, {Fender}, {Waters}, {Dougherty},
  {Koornneef}, {Steele}, \& {van Blokland}}]{Clark:1998}
{Clark}, J.~S., {Fender}, R.~P., {Waters}, L.~B.~F.~M., {et~al.} 1998, \mnras,
  299, L43

\bibitem[{{Clark} {et~al.}(2008){Clark}, {Muno}, {Negueruela}, {Dougherty},
  {Crowther}, {Goodwin}, \& {de Grijs}}]{Clark:2008}
{Clark}, J.~S., {Muno}, M.~P., {Negueruela}, I., {et~al.} 2008, \aap, 477, 147

\bibitem[{{Clark} \& {Negueruela}(2002)}]{Clark:2002}
{Clark}, J.~S. \& {Negueruela}, I. 2002, \aap, 396, L25

\bibitem[{{Clark} {et~al.}(2005){Clark}, {Negueruela}, {Crowther}, \&
  {Goodwin}}]{Clark:2005}
{Clark}, J.~S., {Negueruela}, I., {Crowther}, P.~A., \& {Goodwin}, S.~P. 2005,
  \aap, 434, 949

\bibitem[{{Clark} \& {Porter}(2004)}]{Clark:2004}
{Clark}, J.~S. \& {Porter}, J.~M. 2004, \aap, 427, 839

\bibitem[{{Clark} {et~al.}(2010){Clark}, {Ritchie}, \&
  {Negueruela}}]{Clark:2010}
{Clark}, J.~S., {Ritchie}, B.~W., \& {Negueruela}, I. 2010, \aap, submitted

\bibitem[{{Corcoran}(2008)}]{Corcoran:2008}
{Corcoran}, M.~F. 2008, in Revista Mexicana de Astronomia y Astrofisica
  Conference Series, Vol.~33, 123--128

\bibitem[{{Crowther} {et~al.}(2006){Crowther}, {Hadfield}, {Clark},
  {Negueruela}, \& {Vacca}}]{Crowther:2006}
{Crowther}, P.~A., {Hadfield}, L.~J., {Clark}, J.~S., {Negueruela}, I., \&
  {Vacca}, W.~D. 2006, \mnras, 372, 1407

\bibitem[{{de Jager}(1998)}]{deJager:1998}
{de Jager}, C. 1998, \aapr, 8, 145

\bibitem[{{de Martino} {et~al.}(1992){de Martino}, {Vittone}, {Rossi}, \&
  {Giovannelli}}]{deMartino:1992}
{de Martino}, D., {Vittone}, A.~A., {Rossi}, C., \& {Giovannelli}, F. 1992,
  \aap, 254, 266

\bibitem[{{de Wit} {et~al.}(2008){de Wit}, {Oudmaijer}, {Fujiyoshi}, {Hoare},
  {Honda}, {Kataza}, {Miyata}, {Okamoto}, {Onaka}, {Sako}, \&
  {Yamashita}}]{Dewit:2008}
{de Wit}, W.~J., {Oudmaijer}, R.~D., {Fujiyoshi}, T., {et~al.} 2008, \apjl,
  685, L75

\bibitem[{{Dougherty} {et~al.}(2005){Dougherty}, {Beasley}, {Claussen},
  {Zauderer}, \& {Bolingbroke}}]{Dougherty:2005}
{Dougherty}, S.~M., {Beasley}, A.~J., {Claussen}, M.~J., {Zauderer}, B.~A., \&
  {Bolingbroke}, N.~J. 2005, \apj, 623, 447

\bibitem[{{Dougherty} \& {Williams}(2000)}]{Dougherty:2000}
{Dougherty}, S.~M. \& {Williams}, P.~M. 2000, \mnras, 319, 1005

\bibitem[{{Duncan} \& {White}(2002)}]{Duncan:2002}
{Duncan}, R.~A. \& {White}, S.~M. 2002, \mnras, 330, 63

\bibitem[{{Fullerton} {et~al.}(2006){Fullerton}, {Massa}, \&
  {Prinja}}]{Fullerton:2006}
{Fullerton}, A.~W., {Massa}, D.~L., \& {Prinja}, R.~K. 2006, \apj, 637, 1025

\bibitem[{{Gehrz} {et~al.}(1995){Gehrz}, {Hayward}, {Houck}, {Miles},
  {Hjellming}, {Jones}, {Woodward}, {Prentice}, {Forrest}, {Libonate}, \&
  {Solomon}}]{Gehrz:1995}
{Gehrz}, R.~D., {Hayward}, T.~L., {Houck}, J.~R., {et~al.} 1995, \apj, 439, 417

\bibitem[{{Groh} {et~al.}(2006){Groh}, {Damineli}, {Teodoro}, \&
  {Barbosa}}]{Groh:2006}
{Groh}, J.~H., {Damineli}, A., {Teodoro}, M., \& {Barbosa}, C.~L. 2006, \aap,
  457, 591

\bibitem[{{Groh} {et~al.}(2009){Groh}, {Hillier}, {Damineli}, {Whitelock},
  {Marang}, \& {Rossi}}]{Groh:2009}
{Groh}, J.~H., {Hillier}, D.~J., {Damineli}, A., {et~al.} 2009, \apj, 698, 1698

\bibitem[{{Higgs} {et~al.}(1978){Higgs}, {Feldman}, \&
  {Smolinski}}]{Higgs:1978}
{Higgs}, L.~A., {Feldman}, P.~A., \& {Smolinski}, J. 1978, \apjl, 220, L109

\bibitem[{{Hoare} {et~al.}(2007){Hoare}, {Kurtz}, {Lizano}, {Keto}, \&
  {Hofner}}]{Hoare:2007}
{Hoare}, M.~G., {Kurtz}, S.~E., {Lizano}, S., {Keto}, E., \& {Hofner}, P. 2007,
  in Protostars and Planets V, ed. {B.~Reipurth, D.~Jewitt, \& K.~Keil},
  181--196

\bibitem[{{Ignace} \& {Churchwell}(2004)}]{Ignace:2004}
{Ignace}, R. \& {Churchwell}, E. 2004, \apj, 610, 351

\bibitem[{{Israelian} {et~al.}(1999){Israelian}, {Lobel}, \&
  {Schmidt}}]{Israelian:1999}
{Israelian}, G., {Lobel}, A., \& {Schmidt}, M.~R. 1999, \apjl, 523, L145

\bibitem[{{Jura} \& {Kleinmann}(1990)}]{Jura:1990}
{Jura}, M. \& {Kleinmann}, S.~G. 1990, \apjs, 73, 769

\bibitem[{{Kothes} \& {Dougherty}(2007)}]{Kothes:2007}
{Kothes}, R. \& {Dougherty}, S.~M. 2007, \aap, 468, 993

\bibitem[{{Lang}(2003)}]{Lang:2003}
{Lang}, C.~C. 2003, in IAU Symposium, ed. K.~{van der Hucht}, A.~{Herrero}, \&
  C.~{Esteban}, 497--504

\bibitem[{{Leitherer} {et~al.}(1995){Leitherer}, {Chapman}, \&
  {Koribalski}}]{Leitherer:1995}
{Leitherer}, C., {Chapman}, J.~M., \& {Koribalski}, B. 1995, \apj, 450, 289

\bibitem[{{Leitherer} {et~al.}(1997){Leitherer}, {Chapman}, \&
  {Koribalski}}]{Leitherer:1997}
{Leitherer}, C., {Chapman}, J.~M., \& {Koribalski}, B. 1997, \apj, 481, 898

\bibitem[{{Leitherer} \& {Robert}(1991)}]{Leitherer:1991}
{Leitherer}, C. \& {Robert}, C. 1991, \apj, 377, 629

\bibitem[{{Lobel} {et~al.}(2003){Lobel}, {Dupree}, {Stefanik}, {Torres},
  {Israelian}, {Morrison}, {de Jager}, {Nieuwenhuijzen}, {Ilyin}, \&
  {Musaev}}]{Lobel:2003}
{Lobel}, A., {Dupree}, A.~K., {Stefanik}, R.~P., {et~al.} 2003, \apj, 583, 923

\bibitem[{{Lobel} {et~al.}(1998){Lobel}, {Israelian}, {de Jager}, {Musaev},
  {Parker}, \& {Mavrogiorgou}}]{Lobel:1998}
{Lobel}, A., {Israelian}, G., {de Jager}, C., {et~al.} 1998, \aap, 330, 659

\bibitem[{{Martins} {et~al.}(2007){Martins}, {Genzel}, {Hillier}, {Eisenhauer},
  {Paumard}, {Gillessen}, {Ott}, \& {Trippe}}]{Martins:2007}
{Martins}, F., {Genzel}, R., {Hillier}, D.~J., {et~al.} 2007, \aap, 468, 233

\bibitem[{{Martins} {et~al.}(2005){Martins}, {Schaerer}, {Hillier},
  {Meynadier}, {Heydari-Malayeri}, \& {Walborn}}]{Martins:2005}
{Martins}, F., {Schaerer}, D., {Hillier}, D.~J., {et~al.} 2005, \aap, 441, 735

\bibitem[{{Mengel} \& {Tacconi-Garman}(2007)}]{Mengel:2007}
{Mengel}, S. \& {Tacconi-Garman}, L.~E. 2007, \aap, 466, 151

\bibitem[{{Moffat} {et~al.}(2002){Moffat}, {Corcoran}, {Stevens}, {Skalkowski},
  {Marchenko}, {M{\"u}cke}, {Ptak}, {Koribalski}, {Brenneman}, {Mushotzky},
  {Pittard}, {Pollock}, \& {Brandner}}]{Moffat:2002}
{Moffat}, A.~F.~J., {Corcoran}, M.~F., {Stevens}, I.~R., {et~al.} 2002, \apj,
  573, 191

\bibitem[{{Mokiem} {et~al.}(2007){Mokiem}, {de Koter}, {Vink}, {Puls}, {Evans},
  {Smartt}, {Crowther}, {Herrero}, {Langer}, {Lennon}, {Najarro}, \&
  {Villamariz}}]{Mokiem:2007}
{Mokiem}, M.~R., {de Koter}, A., {Vink}, J.~S., {et~al.} 2007, \aap, 473, 603

\bibitem[{{Monnier} {et~al.}(1999){Monnier}, {Tuthill}, \&
  {Danchi}}]{Monnier:1999}
{Monnier}, J.~D., {Tuthill}, P.~G., \& {Danchi}, W.~C. 1999, \apjl, 525, L97

\bibitem[{{Morris} \& {Jura}(1983)}]{Morris:1983}
{Morris}, M. \& {Jura}, M. 1983, \apj, 267, 179

\bibitem[{{Muno} {et~al.}(2006{\natexlab{a}}){Muno}, {Clark}, {Crowther},
  {Dougherty}, {de Grijs}, {Law}, {McMillan}, {Morris}, {Negueruela}, {Pooley},
  {Portegies Zwart}, \& {Yusef-Zadeh}}]{Muno:2006a}
{Muno}, M.~P., {Clark}, J.~S., {Crowther}, P.~A., {et~al.} 2006{\natexlab{a}},
  \apjl, 636, L41

\bibitem[{{Muno} {et~al.}(2006{\natexlab{b}}){Muno}, {Law}, {Clark},
  {Dougherty}, {de Grijs}, {Portegies Zwart}, \& {Yusef-Zadeh}}]{Muno:2006b}
{Muno}, M.~P., {Law}, C., {Clark}, J.~S., {et~al.} 2006{\natexlab{b}}, \apj,
  650, 203

\bibitem[{{Najarro} {et~al.}(2009){Najarro}, {Figer}, {Hillier}, {Geballe}, \&
  {Kudritzki}}]{Najarro:2009}
{Najarro}, F., {Figer}, D.~F., {Hillier}, D.~J., {Geballe}, T.~R., \&
  {Kudritzki}, R.~P. 2009, \apj, 691, 1816

\bibitem[{{Negueruela} \& {Clark}(2005)}]{Negueruela:2005}
{Negueruela}, I. \& {Clark}, J.~S. 2005, \aap, 436, 541

\bibitem[{{Negueruela} {et~al.}(2010){Negueruela}, {Clark}, \&
  {Ritchie}}]{Negueruela:2010}
{Negueruela}, I., {Clark}, J.~S., \& {Ritchie}, B.~W. 2010, \aap, submitted.

\bibitem[{{Oudmaijer} {et~al.}(2009){Oudmaijer}, {Davies}, {de Wit}, \&
  {Patel}}]{Oudmaijer:2009}
{Oudmaijer}, R.~D., {Davies}, B., {de Wit}, W., \& {Patel}, M. 2009, in
  Astronomical Society of the Pacific Conference Series, Vol. 412, Astronomical
  Society of the Pacific Conference Series, ed. {D.~G.~Luttermoser,
  B.~J.~Smith, \& R.~E.~Stencel}, 17--32

\bibitem[{{Oudmaijer} {et~al.}(1996){Oudmaijer}, {Groenewegen}, {Matthews},
  {Blommaert}, \& {Sahu}}]{Oudmaijer:1996}
{Oudmaijer}, R.~D., {Groenewegen}, M.~A.~T., {Matthews}, H.~E., {Blommaert},
  J.~A.~D.~L., \& {Sahu}, K.~C. 1996, \mnras, 280, 1062

\bibitem[{{Patel} {et~al.}(2008){Patel}, {Oudmaijer}, {Vink}, {Bjorkman},
  {Davies}, {Groenewegen}, {Miroshnichenko}, \& {Mottram}}]{Patel:2008}
{Patel}, M., {Oudmaijer}, R.~D., {Vink}, J.~S., {et~al.} 2008, \mnras, 385, 967

\bibitem[{{Pittard}(2009)}]{Pittard:2009a}
{Pittard}, J.~M. 2009, \mnras, 396, 1743

\bibitem[{{Pittard} {et~al.}(2006){Pittard}, {Dougherty}, {Coker}, {O'Connor},
  \& {Bolingbroke}}]{Pittard:2006}
{Pittard}, J.~M., {Dougherty}, S.~M., {Coker}, R.~F., {O'Connor}, E., \&
  {Bolingbroke}, N.~J. 2006, \aap, 446, 1001

\bibitem[{{Pittard} \& {Parkin}(2009)}]{Pittard:2009b}
{Pittard}, J.~M. \& {Parkin}, E.~R. 2009, ArXiv e-prints

\bibitem[{{Povich} {et~al.}(2007){Povich}, {Stone}, {Churchwell}, {Zweibel},
  {Wolfire}, {Babler}, {Indebetouw}, {Meade}, \& {Whitney}}]{Povich:2007}
{Povich}, M.~S., {Stone}, J.~M., {Churchwell}, E., {et~al.} 2007, \apj, 660,
  346

\bibitem[{{Reid}(2006)}]{Reid:2006}
{Reid}, R.~I. 2006, \mnras, 367, 1766

\bibitem[{{Repolust} {et~al.}(2004){Repolust}, {Puls}, \&
  {Herrero}}]{Repolust:2004}
{Repolust}, T., {Puls}, J., \& {Herrero}, A. 2004, \aap, 415, 349

\bibitem[{{Ritchie} {et~al.}(2010){Ritchie}, {Clark}, \&
  {Negueruela}}]{Ritchie:2010}
{Ritchie}, B.~W., {Clark}, J.~S., \& {Negueruela}, I. 2010, A\&A, submitted.

\bibitem[{{Ritchie} {et~al.}(2009{\natexlab{a}}){Ritchie}, {Clark},
  {Negueruela}, \& {Crowther}}]{Ritchie:2009a}
{Ritchie}, B.~W., {Clark}, J.~S., {Negueruela}, I., \& {Crowther}, P.~A.
  2009{\natexlab{a}}, \aap, 507, 1585

\bibitem[{{Ritchie} {et~al.}(2009{\natexlab{b}}){Ritchie}, {Clark},
  {Negueruela}, \& {Najarro}}]{Ritchie:2009b}
{Ritchie}, B.~W., {Clark}, J.~S., {Negueruela}, I., \& {Najarro}, F.
  2009{\natexlab{b}}, \aap, 507, 1597

\bibitem[{{Rybicki} \& {Lightman}(1986)}]{Rybicki:1986}
{Rybicki}, G.~B. \& {Lightman}, A.~P. 1986, {Radiative Processes in
  Astrophysics}, ed. G.~B. {Rybicki} \& A.~P. {Lightman}

\bibitem[{{Sault} {et~al.}(1995){Sault}, {Teuben}, \& {Wright}}]{Sault:1995}
{Sault}, R.~J., {Teuben}, P.~J., \& {Wright}, M.~C.~H. 1995, in ASP Conf. Ser.
  77: Astronomical Data Analysis Software and Systems IV, 433--436

\bibitem[{{Schuster} {et~al.}(2006){Schuster}, {Humphreys}, \&
  {Marengo}}]{Schuster:2006}
{Schuster}, M.~T., {Humphreys}, R.~M., \& {Marengo}, M. 2006, \aj, 131, 603

\bibitem[{{Serabyn} {et~al.}(1991){Serabyn}, {Lacy}, \&
  {Achtermann}}]{Serabyn:1991}
{Serabyn}, E., {Lacy}, J.~H., \& {Achtermann}, J.~M. 1991, \apj, 378, 557

\bibitem[{{Setia Gunawan} {et~al.}(2003){Setia Gunawan}, {Chapman}, {Stevens},
  {Rauw}, \& {Leitherer}}]{SetiaGunawan:2003}
{Setia Gunawan}, D.~Y.~A., {Chapman}, J.~M., {Stevens}, I.~R., {Rauw}, G., \&
  {Leitherer}, C. 2003, in IAU Symposium, ed. K.~{van der Hucht}, A.~{Herrero},
  \& C.~{Esteban}, 230--231

\bibitem[{{Shepherd}(1997)}]{Shepherd:1997}
{Shepherd}, M.~C. 1997, in ASP Conf. Ser. 125: Astronomical Data Analysis
  Software and Systems VI, 77--84

\bibitem[{{Skinner} {et~al.}(2006){Skinner}, {Simmons}, {Zhekov}, {Teodoro},
  {Damineli}, \& {Palla}}]{Skinner:2006}
{Skinner}, S.~L., {Simmons}, A.~E., {Zhekov}, S.~A., {et~al.} 2006, \apjl, 639,
  L35

\bibitem[{{Smith} {et~al.}(2001){Smith}, {Humphreys}, {Davidson}, {Gehrz},
  {Schuster}, \& {Krautter}}]{Smith:2001}
{Smith}, N., {Humphreys}, R.~M., {Davidson}, K., {et~al.} 2001, \aj, 121, 1111

\bibitem[{{Smith} \& {Owocki}(2006)}]{Smith:2006}
{Smith}, N. \& {Owocki}, S.~P. 2006, \apjl, 645, L45

\bibitem[{{Stahl} {et~al.}(1991){Stahl}, {Wolf}, {Aab}, \&
  {Smolinski}}]{Stahl:1991}
{Stahl}, O., {Wolf}, B., {Aab}, O., \& {Smolinski}, J. 1991, \aap, 252, 693

\bibitem[{{Stevens} {et~al.}(1992){Stevens}, {Blondin}, \&
  {Pollock}}]{Stevens:1992}
{Stevens}, I.~R., {Blondin}, J.~M., \& {Pollock}, A.~M.~T. 1992, \apj, 386, 265

\bibitem[{{Stickland} \& {Harmer}(1978)}]{Stickland:1978}
{Stickland}, D.~J. \& {Harmer}, D.~L. 1978, \aap, 70, L53

\bibitem[{{Sylvester} {et~al.}(1998){Sylvester}, {Skinner}, \&
  {Barlow}}]{Sylvester:1998}
{Sylvester}, R.~J., {Skinner}, C.~J., \& {Barlow}, M.~J. 1998, \mnras, 301,
  1083

\bibitem[{{Tafoya} {et~al.}(2004){Tafoya}, {G{\'o}mez}, \&
  {Rodr{\'{\i}}guez}}]{Tafoya:2004}
{Tafoya}, D., {G{\'o}mez}, Y., \& {Rodr{\'{\i}}guez}, L.~F. 2004, \apj, 610,
  827

\bibitem[{{Taylor} {et~al.}(1987){Taylor}, {Pottasch}, \&
  {Zhang}}]{Taylor:1987}
{Taylor}, A.~R., {Pottasch}, S.~R., \& {Zhang}, C.~Y. 1987, \aap, 171, 178

\bibitem[{{Tuthill} {et~al.}(1999){Tuthill}, {Monnier}, \&
  {Danchi}}]{Tuthill:1999}
{Tuthill}, P.~G., {Monnier}, J.~D., \& {Danchi}, W.~C. 1999, \nat, 398, 487

\bibitem[{{van der Hucht} {et~al.}(1997){van der Hucht}, {Schrijver},
  {Stenholm}, {Lundstrom}, {Moffat}, {Marchenko}, {Seggewiss}, {Setia Gunawan},
  {Sutantyo}, {van den Heuvel}, {de Cuyper}, \& {Gomez}}]{vanderHucht:1997}
{van der Hucht}, K.~A., {Schrijver}, H., {Stenholm}, B., {et~al.} 1997, New
  Astronomy, 2, 245

\bibitem[{{van Loo} {et~al.}(2006){van Loo}, {Runacres}, \&
  {Blomme}}]{vanloo:2006}
{van Loo}, S., {Runacres}, M.~C., \& {Blomme}, R. 2006, \aap, 452, 1011

\bibitem[{{Voors} {et~al.}(2000){Voors}, {Waters}, {de Koter}, {Bouwman},
  {Morris}, {Barlow}, {Sylvester}, {Trams}, \& {Lamers}}]{Voors:2000}
{Voors}, R.~H.~M., {Waters}, L.~B.~F.~M., {de Koter}, A., {et~al.} 2000, \aap,
  356, 501

\bibitem[{{Watson} {et~al.}(2008){Watson}, {Povich}, {Churchwell}, {Babler},
  {Chunev}, {Hoare}, {Indebetouw}, {Meade}, {Robitaille}, \&
  {Whitney}}]{Watson:2008}
{Watson}, C., {Povich}, M.~S., {Churchwell}, E.~B., {et~al.} 2008, \apj, 681,
  1341

\bibitem[{{Westerlund}(1961)}]{Westerlund:1961}
{Westerlund}, B. 1961, \pasp, 73, 51

\bibitem[{{Westerlund}(1987)}]{Westerlund:1987}
{Westerlund}, B.~E. 1987, \aaps, 70, 311

\bibitem[{{Westmoquette} {et~al.}(2007){Westmoquette}, {Exter}, {Smith}, \&
  {Gallagher}}]{Westmoquette:2007}
{Westmoquette}, M.~S., {Exter}, K.~M., {Smith}, L.~J., \& {Gallagher}, J.~S.
  2007, \mnras, 381, 894

\bibitem[{{Williams}(1999)}]{Williams:1999}
{Williams}, P.~M. 1999, in IAU Symp. 193: Wolf-Rayet Phenomena in Massive Stars
  and Starburst Galaxies, ed. K.~A. {van der Hucht}, G.~{Koenigsberger}, \&
  P.~R.~J. {Eenens}, 267--277

\bibitem[{{Williams} {et~al.}(1997){Williams}, {Dougherty}, {Davis}, {van der
  Hucht}, {Bode}, \& {Setia Gunawan}}]{Williams:1997}
{Williams}, P.~M., {Dougherty}, S.~M., {Davis}, R.~J., {et~al.} 1997, \mnras,
  289, 10

\bibitem[{{Williams} {et~al.}(2009){Williams}, {Marchenko}, {Marston},
  {Moffat}, {Varricatt}, {Dougherty}, {Kidger}, {Morbidelli}, \&
  {Tapia}}]{Williams:2009}
{Williams}, P.~M., {Marchenko}, S.~V., {Marston}, A.~P., {et~al.} 2009, \mnras,
  395, 1749

\bibitem[{{Yusef-Zadeh} \& {Morris}(1991)}]{Yusef-Zadeh:1991}
{Yusef-Zadeh}, F. \& {Morris}, M. 1991, \apjl, 371, L59

\end{thebibliography}

\appendix
\section{Modelling radio emission from a circumstellar envelope}
\label{sec:appendix}
\begin{figure}[t]
\vspace{7.3cm}
\includegraphics{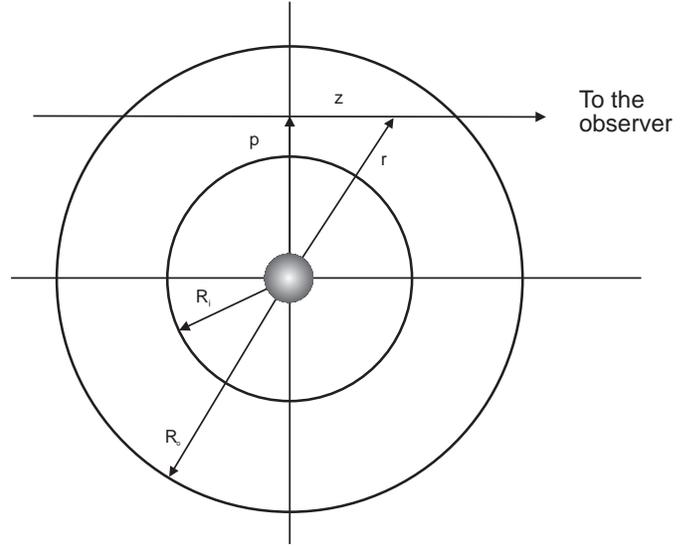}
\caption[]{Geometry of a model circumstellar envelope with outer
radius $R_o$ and inner radius $R_i$. For $R_i <r<R_o$ the plasma
density $\propto r^{-2}$ and is zero for $r<R_i$.}
\label{fig:modelgeometry}
\end{figure}

For a spherically symmetric isothermal plasma envelope of temperature
$T$ and outer radius $R_o$ at distance $D$ from the Sun, the total
flux at frequency $\nu$ is given by
\begin{equation}
F_\nu ={{2 \pi}\over {D^2}}\bigl ({{2 k T \nu^2}\over{c^2}} \bigr )
\int_0^{R_o} (1-e^{-\tau_\nu(p)})p\, dp\,, 
\label{eqn:flux}
\end{equation}
where $\tau_\nu(p)$ is the optical depth along a line-of-sight at
distance $p$ from the line-of-sight to the centre of the
envelope\footnote{$p$ is often referred to as the ``impact
parameter''.}  (Fig.~\ref{fig:modelgeometry}) For a Maxwellian plasma
in LTE the opacity is given by
\begin{displaymath}
\tau_\nu = 3.6\times10^{-2}T_e^{-3/2}\nu^{-2} g_\nu \gamma Z^2 \int_0^\infty n_e^2 dz\,,
\end{displaymath}
for electron temperature $T_e$, $g_\nu$ the gaunt free-free factor
\citep[see e.g.][]{Leitherer:1991}, $Z$ the rms charge of the plasma ,
$\gamma$ the ratio of electron to ions, and $n_e$ is the electron
density along a line-of-sight segment length $dz$ \citep{Rybicki:1986}.

For a steady-state, constant velocity stellar wind of mass-loss rate
$\dot M$ and terminal velocity v$_\infty$, the equation of continuity
requires $n_e {\rm v}_\infty r^2 =$~const, leading to
\begin{displaymath}
n_e = {{\dot M}\over {4 \pi {\rm v}_\infty \mu {\rm m}_H}} \bigl ({1\over r^2}\bigr )\,,
\end{displaymath}
with $\mu$ the atomic weight of the plasma, and m$_H$ the mass of
hydrogen. The free-free opacity is then given by
\begin{equation}
\tau_\nu(p) = A \int_0^\infty {{1}\over{(p^2 + z^2)^2}}\, dz = A\,I_2(p)
\label{eqn:opacity}
\end{equation}
where 
\begin{displaymath}
A=3.6\times10^{-2}T_e^{-3/2}\nu^{-2} g_\nu \gamma \bigl ({{\dot
M \, Z}\over {4 \pi {\rm v}_\infty \mu {\rm m}_H}}\bigr )^2.
\end{displaymath}
Analytic solutions for the integral $I_2(p)$ in Eqn.~\ref{eqn:opacity} and
the envelope geometry shown in Fig.~\ref{fig:modelgeometry} can
be derived, namely:
\begin{eqnarray}
\label{eqn:opacity_solutions1}
I_2(p)
&=& \frac{1}{2p^2} \left [\frac{(R_o^2 - p^2)^{\frac{1}{2}}}{R_o^2}
  - \frac{(R_i^2-p^2)^{\frac{1}{2}}}{R_i^2}\right ] \nonumber\\
&&+\frac{1}{2p^3}
\left [\tan^{-1}\left (\frac{(R_o^2-p^2)^{\frac{1}{2}}}{p}\right ) 
-      \tan^{-1}\left (\frac{(R_i^2-p^2)^{\frac{1}{2}}}{p}\right )\right ] 
\end{eqnarray}
if $0 < p < R_i$ and 
\begin{eqnarray}
\label{eqn:opacity_solutions2}
I_2(p)
&=&\frac{1}{2p^3}\left [\frac{p(R_o^2-p^2)^{\frac{1}{2}}}{R_o^2} +
      \tan^{-1}\left (\frac{(R_o^2-p^2)^{\frac{1}{2}}}{p}\right )\right ]
\end{eqnarray}
if $R_i < p < R_o$.

Determining the flux from an extended nebula at any given frequency
requires using Eqns.~\ref{eqn:opacity} and the two conditional cases
for $I_2)p)$ given by Eqn.~\ref{eqn:opacity_solutions1}
and~\ref{eqn:opacity_solutions2} to determine the opacity along a
particular line-of-sight, before solving Eqn.~\ref{eqn:flux}. To model
the flux from a ``standard'' stellar wind, $R_i$ is set to the radius
of the underlying star $R_*$ (for the models considered here $R_* <<
R_o$, so it is assumed that $R_i\equiv 0$) and only the conditional
case for $I_2(p)$ given in Eqn.~\ref{eqn:opacity_solutions2} needs to
be considered.

\end{document}